\documentclass[12pt, prd,nofootinbib,preprint,superscriptaddress]{revtex4}

\usepackage{amsmath, amssymb, amsthm, mathtools,graphicx, epsfig, fancyhdr,epsfig, slashed, mathrsfs,physics,empheq
,xstring,babel,multirow}
\usepackage[utf8]{inputenc}
\usepackage{booktabs}

\usepackage{tikzsymbols}
\usepackage{tikz,xcolor
}
\usepackage{natbib}
\usepackage{float}

\definecolor{cadmiumred}{rgb}{0.89, 0.0, 0.13}
\definecolor{candyapplered}{rgb}{1.0, 0.03, 0.0}
\definecolor{copper}{rgb}{0.72, 0.45, 0.2}
\definecolor{darkmagenta}{rgb}{0.55, 0.0, 0.55}

\usepackage{hyperref}
\hypersetup{colorlinks,linkcolor={blue}
,citecolor={cyan},urlcolor={dukeblue}}

\definecolor{lime}{HTML}{A6CE39}
\DeclareRobustCommand{\orcidicon}{
	\begin{tikzpicture}
	\draw[lime, fill=lime] (0,0) 
	circle [radius=0.2] 
	node[white] {{\fontfamily{qag}\selectfont \tiny ID}};
	\draw[white, fill=white] (-0.0625,0.095) 
	circle [radius=0.007];
	\end{tikzpicture}
	\hspace{-2mm}
}

\foreach \x in {Anish,Lalak, Shila,Supratik}{\expandafter\xdef\csname orcid\x\endcsname{\noexpand\href{https://orcid.org/\csname orcidauthor\x\endcsname}
			{\noexpand\orcidicon}}
}

\newcommand{\be}{\begin{equation}}
\newcommand{\ee}{\end{equation}}
\newcommand{\bea}{\begin{eqnarray}}
\newcommand{\eea}{\end{eqnarray}}
\newcommand{\bc}{\begin{cases}}
\newcommand{\ec}{\end{cases}}

\definecolor{dukeblue}{rgb}{0.0, 0.0, 0.61}

\def\hds#1{\href{https://doi.org/#1}}

 \usepackage{
 soul,
 cancel,
 enumitem,
 multirow,
 slashed,
 upgreek, 
 physics, cases}
 \usepackage{easyReview}
\usepackage{bbold}
\usepackage[normalem]{ulem}

\newcommand{\eq}[1]{\begin{align}#1\end{align}}

\definecolor{amber(sae/ece)}{rgb}{1.0, 0.49, 0.0}

\definecolor{blue(ncs)}{rgb}{0.0, 0.53, 0.74}

\definecolor{darkviolet}{rgb}{0.58, 0.0, 0.83}

\definecolor{americanrose}{rgb}{1.0, 0.01, 0.24}

\definecolor{jazzberryjam}{rgb}{0.65, 0.04, 0.37}
\definecolor{jonquil}{rgb}{0.98, 0.85, 0.37}
\definecolor{mikadoyellow}{rgb}{1.0, 0.77, 0.05}
\definecolor{schoolbusyellow}{rgb}{1.0, 0.85, 0.0}

\newcommand{\ba}{\begin{eqnarray}}
\newcommand{\ea}{\end{eqnarray}}

\newcommand{\unit}[1]{\, \text{#1}}

\newcommand{\lt}{\left }
\newcommand{\rt}{\right }
\newcommand{\seq}{\simeq}
\newcommand{\gsim}{\gtrsim}
\newcommand{\lsim}{\lesssim}

\def\gs{g_{\star}}
\def\gss{g_{\star, s}}
\def\mpl{M_P}
\def\GeV{\unit{GeV}}
\def\MeV{\unit{MeV}}
\def\keV{\unit{keV}}

\def\cmb{\text{CMB}}
\def\lep{\rm lep}

\def\ncmb{N_{\cmb}}

\newcommand{\Planck}{\textit{Planck}}

\newcommand{\KeckArray}{\textit{Keck Array}}
\newcommand{\BICEP}{\textsc{Bicep}}

\newcommand{\cmbsfour}{{CMB-S4}}
\newcommand{\LB}{\textit{LiteBIRD}}

\def\Trh{T_{\rm rh}}
\def\rh{{\rm rh}}
\def\rr{\r_{\rm rad}}
\def\re{{\rm re}}
\def\qur{{\rm qur}}
\def\qud{{\rm qud}}

\def\Br{\text{Br}}
\def\cc{\bar{\chi}\chi}

\def\baremass{{m_b}}

\def\lop{{\rm 1-loop}}

\def\sm{\text{SM}}
\def\bsm{\text{BSM}}
\def\dm{\text{DM}}
\def\cdm{CDM}
\def\bbn{BBN}

\def\ncmb{{\cal N}_{\cmb}}

\def\td{\text{d}}

\def\r{\rho}

\def\G{\Gamma}
\def\L{\Lambda}
\def\vp{\varphi}

\def\mmoon{(m)BM3}
\def\msun{(m)BM7}
\def\Pmoon{(P)BM3}
\def\Psun{(P)BM7}
\def\nmcw{\text{NMCW}}
\def\nmpq{\text{NMPQ}}

\def\lop{{\rm 1-loop}}

\def\cs{\boldsymbol{a}}
\def\csn{\boldsymbol{\mathcal{A}}}
\def\hubble{{\cal H}}

\def\cc{\bar{\chi}\chi}
\def\lH{\lambda_H}

\def\mH{m_H}
\def\mc{m_{\chi}}
\def\yc{y_\chi}
\def\Yc{Y_\chi}

\def\lO{\lambda_{12}}

\def\lT{\lambda_{22}}

\def\HH{H^\dagger H}

\def\MN{{M_N}}
\def\Yu{{\cal Y}}
\def\Trh{T_{\rm rh}}
\def\Tmax{T_{max}}
\def\rad{\text{rad}}
\def\Br{\text{Br}}

\def\cq{{\cal Q}}

\def\jframe{\text{Jordan frame}}
\def\eframe{\text{Einstein frame}}

\usepackage[capitalise]{cleveref}
\Crefname{figure}{Fig.}{Figs.}
\Crefname{section}{Sec.}{Secs.}
\newcommand{\Ccite}[1]{%
\IfSubStr{#1}{,}{Refs.~}{Ref.~}\cite{#1}%
}

\begin{document}


\title{Post-inflationary Leptogenesis and Dark Matter production: \it{Metric versus Palatini formalism}}

\author{Anish Ghoshal\orcidAnish{}}
\email{anish.ghoshal@fuw.edu.pl}
\author{Zygmunt Lalak\orcidLalak{}}
\email{zygmunt.lalak@fuw.edu.pl}
\affiliation{Institute of Theoretical Physics, Faculty of Physics, University of Warsaw, ul. Pasteura 5, 02-093 Warsaw, Poland}

\author{Supratik Pal\orcidSupratik{}}
\email{supratik@isical.ac.in}
\affiliation{Physics and Applied Mathematics Unit, Indian Statistical Institute, 203 B.T. Road, Kolkata 700108, India}
\affiliation{Technology Innovation Hub on Data Science, Big Data Analytics and Data Curation, Indian Statistical Institute, 203 B.T. Road, Kolkata 700108, India}

\author{Shiladitya Porey\orcidShila{}}
\email{shiladityamailbox@gmail.com}
\affiliation{Department of Physics, Novosibirsk State University, Pirogova 2, 630090 Novosibirsk, Russia}

\begin{abstract}
{\it
We investigate production of non-thermal dark matter particles and heavy sterile neutrinos from inflaton during the reheating era, which is preceded by a slow-roll inflationary epoch with a quartic potential and non-minimal coupling ($\xi$) between inflaton and gravity. We compare our analysis between metric and Palatini formalism. For the latter, the tensor-to-scalar ratio, r, decreases with $\xi$. We find that for $\xi=0.5$ and number of $e$-folds $\sim 60$, $r$ can be as small as $\sim {\cal O}\left(10^{-3}\right)$ which may be validated at future reaches of upcoming CMB observation such as \cmbsfour~etc. We identify the permissible range of Yukawa coupling $y_\chi$ between inflaton and fermionic DM $\chi$, to be ${\cal O}\left(10^{-3.5}\right)\gtrsim y_\chi \gtrsim {\cal O}\left(10^{-20}\right)$ for metric formalism and ${\cal O}\left(10^{-4}\right)\gtrsim y_\chi \gtrsim {\cal O}\left(10^{-11}\right)$ for Palatini formalism which is consistent with current PLANCK data and also within the reach of future CMB experiments. For the scenario of leptogenesis via the decay of sterile neutrinos produced from inflaton decay, we also investigate the parameter space involving heavy neutrino mass $M_{N_1}$ and Yukawa coupling $y_{N_1}$ of sterile neutrino with inflaton, which are consistent with current CMB data and successful generation of the observed baryon asymmetry of the universe via leptogenesis. In contrast to metric formalism, in the case of Palatini formalism, for successful leptogenesis to occur, we find that $y_{N_1}$ has a very narrow allowable range and is severely constrained from the consistency with CMB predictions. 
}
\end{abstract}

\maketitle

\section{Introduction}





The strongest evidences that upholds the standard model of cosmology comes from remarkable agreement between the theoretical prediction and observed data about Big Bang Nucleosynthesis (\bbn), and the detection of CMB photons in the present universe. During the BBN era, which occurs when the temperature of the universe
is between $\sim 1 \MeV$ and $\sim 10 \keV$, light elements e.g. deuterium, helium-3,
helium-4, and lithium-7 are formed~\cite{Bambi:2015mba}. The production mechanism for these elements during BBN is based on well established theory of standard model (\sm) of particle physics. Consequently, the abundance of those light elements put stringent constraints on BSM physics.

Conversely, numerical analysis of data obtained from \cmb~also places constraints on cosmological parameters such as present-day energy density of baryons and dark matter (DM), among others. Almost all elementary particles in the \sm~of particle physics that exist in our universe have corresponding antiparticles with the same mass and lifetime but with opposite charges. %
All those \sm~particles were relativistic in the early universe, when the temperature was higher than the mass of the top-quark. %
We assume their chemical potential was negligible at that time so they existed likely in local thermal equilibrium with the SM photons, which suggests 
that the numbers of particles and antiparticles were initially equal. %
Therefore, the matter-antimatter asymmetry observed in our present observable universe remains unexplainable. The ratio of the number density of baryons to photons in the present universe obtained from \cmb~is $\eta_B \sim 6.14\times 10^{-10}$~\cite{Planck:2018vyg,ParticleDataGroup:2020ssz,Fields:2019pfx}. This result is consistent with \bbn~analysis and observed small fraction of antiproton (${\cal O}\qty(10^{-4})$) and anti-helium-4 (${\cal O}\qty(10^{-7})$) from cosmic rays~\cite{Bambi:2015mba}, and the ratio of antibaryon to baryon obtained from the two colliding clusters of galaxies in Bullet Cluster~\cite{Steigman:2008ap}. Moreover, 
if there is a nearby antimatter domain, it must not be located inside the cosmological horizon. Consequently, a universe characterized by matter-antimatter symmetry is empirically ruled out~\cite{Cohen:1997ac}. The predominance of matter over antimatter in the early universe can be generated through a dynamic physical process if that violates (i) baryon number B, (ii) C- and CP-invariance, and (iii) thermal equilibrium~\cite{Sakharov:1967dj}. These conditions are known as {\it Sakharov} principles. Although, numerous baryogenesis mechanisms have been proposed, including decay of heavy particles~\cite{Sakharov:1967dj}, via evaporation of primordial black holes~\cite{Hawking:1974rv,Zeldovich:1976vw}, the most extensively studied ones being electroweak baryogenesis~\cite{Kuzmin:1985mm,Shaposhnikov:1986jp,Shaposhnikov:1987tw} and baryogenesis through leptogenesis~\cite{Fukugita:1986hr}. Among them the latter involves the decay of heavy right-handed neutrinos, introduced as an extension to the \sm~to achieve light neutrino masses via the Type-I seesaw mechanism~\cite{Minkowski:1977sc,Gell-Mann:1979vob,Yanagida:1979as,Mohapatra:1979ia,Mohapatra:2004zh}, and also addresses the puzzle of tiny neutrino mass generation which is evident from several neutrino oscillation experiments.

Furthermore, CMB data, along with by several other independent cosmological observations,  reveals that cold dark matter (CDM) contributes to $\sim 26\%$ of total mass-energy density of the universe~\cite{Planck:2018vyg,ParticleDataGroup:2022pth}. This contribution is almost five times that of visible matter. 
The particle nature of dark matter remains an enigma to date, as no direct detection of DM experiment found any evidence, including the most popular candidate for particle DM - Weakly Interacting Massive Particles (WIMPs). It is assumed that WIMP particles were in thermal equilibrium in the early universe, and then they decoupled at a later time. Several unsuccessful attempts to detect such particles through the scattering of atomic nuclei or electrons, or by detecting the products of their decay from cosmic rays~\cite{Bergstrom:2013jra,Cuoco:2017iax}, or in particle detectors at CERN~\cite{ATLAS:2022rxn,Pedro:2023zwd,Giagu:2019fmp}, brings forth the alternative assumptions that DM particles are so much feebly interacting that they could never reach in thermal equilibrium in the early universe. They could be produced from the decay of massive particles in the early universe, and unlike thermal DM, their relic density depends on the specific production channels. 


Observation of the temperature of CMB photons over the full sky reveals that temperature anisotropy is even less than $10^{-5}$ in contrast to the expectation of the last scattering surface being $\sim 10^5$ causally disconnected patches~\cite{Kamionkowski:1999qc}. This implies that the largest length scale that enters the causal Horizon today must have existed well inside the earliest epoch. The simplest way to envision such a scenario is by proposing that there was an early epoch during which the energy density of a slowly rolling scalar field, $\phi$, referred to as inflaton, along the slope of its potential dominated the universe. This minimal modification to the standard model of cosmology effectively accounts for the generation of primordial perturbation and flatness of our universe. Although further validation of the inflationary paradigm is still on the cards, this is widely accepted as the {\it new standard model of cosmology} within the scientific community. Despite its stupendous success in overcoming the limitations of Big Bang cosmology and providing the required amount of seeds of matter perturbations, the minimal models of chaotic inflation involving potential $\propto \phi^p$, where $p$ is an integer number, appear to be disfavored by the latest data released by \Planck~mission~\cite{Planck:2018vyg} and data from \BICEP+\KeckArray~\cite{BICEP:2021xfz} regarding unobserved B-mode of primordial tensor perturbations from inflaton. However, as revealed in subsequent studies~\cite{Boubekeur:2015xza,Bezrukov:2008dt,Kallosh:2013pby}, the scenario can be reinstated, bringing these models back to the limelight, by considering that the inflaton is non-minimally coupled to gravity. Such a coupling can be generated from quantum correction~\cite{Hertzberg:2010dc}, and provides a plateau-like region for the inflation to occur. 

An interesting possibility arises when the inflaton is non-minimally coupled to gravity, the predictions of slow-roll inflation, such as the scalar spectral index ($n_s$) and tensor-to-scalar ratio ($r$), may vary between the metric and Palatini formalisms (see~\Ccite{Capozziello:2010ih,Kallosh:2013tua,Jarv:2017azx,Racioppi:2017spw}). For small or near-zero values of non-minimal coupling yield nearly identical predictions in both formalisms, but larger values lead to differences. In the metric formalism, for an inflaton potential $V(\phi)\propto\phi^p$, $n_s$ and $r$ remain independent of non-minimal coupling at the leading order. Conversely, in the Palatini formalism, the predicted values of $r$ vary inversely with non-minimal coupling~\cite{Cheong:2021kyc}. 

%
In metric formalism, space-time metric ($g_{\alpha \beta}$) and its first derivative are the independent variables, whereas in Palatini formalism, space-time metric and metric connection (or, affine connection which is denoted by $\Gamma^\lambda_{\alpha \beta}$) are independent variables~\cite{Bostan:2022swq,Racioppi:2019jsp}. In differential geometry, metric connection measures the intrinsic curvature of the manifold. In metric formalism, it is assumed that $\Gamma^\lambda_{\alpha \beta}=\Gamma^\lambda_{\beta\alpha}$ and $\nabla_\lambda g_{\alpha \beta}=0$, and as a result affine connection is replaced by Levi-Civita connection, $\bar{\Gamma}_{\alpha \beta}^\lambda$, which is defined as~\cite{Bauer:2008zj} 
$\bar{\Gamma}_{\alpha \beta}^\lambda= \frac12 {g^{JF}}^{\lambda \zeta} \lt[\partial_\alpha \lt( {g^{JF}}_{\beta \zeta} \rt) + \partial_\beta \lt( {g^{JF}}_{\alpha \zeta} \rt) -\partial_\zeta \lt( {g^{JF}}_{\alpha\beta} \rt)   \rt]
.$~
In the theory of general relativity, the preference for the Levi-Civita connection arises not only because the torsion part of the connection disappears, but also due to the Equivalence Principle and the importance of aligning affine geodesics and metric geodesics to uphold causality~\cite{Borunda:2008kf}. 
Consequently, in metric formalism, Riemann curvature tensor depends on the second derivative of the space-time metric. Conversely, in Palatini formalism, Riemann curvature tensor depends on the first derivative of the space-time metric~\cite{Bauer:2008zj}, and Ricci tensor depends on connection but independent of the metric tensor~\cite{Borunda:2008kf}. Moreover, in the Palatini formulation, increasing $\xi$ values reduces the value of $r$ significantly ($r\sim {\ncmb}^{-2}$ in metric formalism, and $r\sim \lt(\xi\,\ncmb\rt)^{-2}$ in Palatini formalism, with $\ncmb$ denoting the number of $e$-folds of inflation~\cite{Shaposhnikov:2020fdv}). This provides a potential pathway for rescuing inflationary models which have been ruled out by CMB observations for predicting large values of $r$ within the metric formalism~\cite{Racioppi:2017spw,Artymowski:2016dlz,Kannike:2015kda,Barrie:2016rnv}.
The end of inflation is followed by reheating era that acts as a bridge to transform the inflaton dominated cold universe into our familiar universe dominated by relativistic standard model particles~\cite{Allahverdi:2010xz}.

In this article, we plan to investigate post-inflationary dark matter production and leptogenesis parallelly within these two formalisms. Inflaton can also generate stable but light particles contributing to dark radiation~\cite{Ghoshal:2023phi}. In this study, our focus is on a scenario that involves both the inflaton and a gauge singlet BSM field which if stable is the CDM candidate or can be the unstable right-handed neutrino which decays to SM particles leading to baryogenesis via leptogenesis.  For a list of models incorporating inflation and DM, see~\Ccite{Ghoshal:2022jeo} and references therein. In our work, we assume that \bsm~vectorlike fermionic particles are produced during the reheating era through the decay of inflaton. These particles are assumed stable, non-relativistic, but feebly interacting with the SM relativistic plasma, and contributes to the total CDM density of the universe. We also consider another possibility when a sterile neutrino is produced from the decay of inflaton. Their decay is out-of-equilibrium, C and CP violating process (with 1-loop effects) and violates lepton number, thereby satisfying Sakharov's conditions~\cite{Sakharov:1967dj}. Then, baryon-lepton number violating sphaleron processes coverts lepton asymmetry to baryon asymmetry, see.~\Ccite{Asaka:1999yd,Co:2022bgh,Senoguz:2004ky,Asaka:1999jb,Barrie:2021mwi} 


\textit{This paper is organized as follows:} in~\cref{Sec:Lagrangian}, we discuss the Lagrangian density of inflation non-minimally coupled to gravity, as well as the interaction Lagrangian density during reheating. In~\cref{sec:Quartic slow roll}, we study the benchmark for slow roll inflationary scenario with a quartic potential of inflaton. In~\cref{sec:stability}, we present the stability analysis. In~\cref{sec:reheating}, we focus on reheating and production of non-thermal BSM particles as CDM. In~\cref{sec:lepto} we explore leptogenesis from the decay of sterile neutrino. 

In this work, we use $\hbar=c=k_B=1$ and the value of reduced Planck constant $\mpl=2.4\times 10^{18}\GeV$. In addition, we assume that the space-time metric is diagonal with signature $(+,-,-,-)$.


\section{Lagrangian Density}
\label{Sec:Lagrangian}
%
%

In this article, our focus is on the reheating era following a slow roll inflation, where we consider the non-minimal coupling between the inflaton $\vp$ and the curvature scalar ${\cal R}$, and the production of a fermionic \bsm~particle alongside the SM Higgs particle $h$ during reheating. Then, the action for inflation in Jordan frame is given by%
%
%
%
~\cite{Bostan:2019fvk,Markkanen:2017tun}
\eq{\label{Eq:Jordan frame action}
{\cal S}^{JF}\supset \int \td^4 x  \sqrt{-g^{JF}} \lt(\frac{1}{2}M_P^2 \, \Omega^2(\varphi)\, \mathscr{R}^{JF}   + \frac{1}{2} \lt( \partial \vp\rt)^2  -  V^{JF}(\vp)\rt)\,.
}
We use $JF$ in superscript to refer that the corresponding quantity is defined in Jordan frame. Therefore, $g^{JF}$, $\mathscr{R}^{JF}$, and $V^{JF}(\vp)$ are the determinant of the space-time metric, Ricci scalar and potential of the inflaton in Jordan frame, respectively. Additionally, we consider the form of $\Omega^2(\varphi)\equiv \Omega^2\neq 0$ as%
\footnote{
To ensure each term in the parentheses on the right-hand side of~\cref{Eq:Jordan frame action} has dimension-$4$ with dimensionless coefficients, a common choice for non-minimal coupling is $\Omega^2=1+\xi \vp^2/\mpl^2$.  Despite the absence of interaction between the inflaton and gravity sector at the tree level, non-minimal coupling can emerge at one-loop order. Alternative forms of non-minimal couplings, such as periodic forms (see~\Ccite{Ferreira:2018nav,Ghoshal:2023jvf}), exist or we can include additional covariant scalar terms like $\mathscr{R}^{\mu\nu}\partial_{\mu}\vp\partial_\nu \vp$, where $\mathscr{R}_{\mu\nu}$ is Ricci curvature tensor, within the Lagrangian density. However, limiting considerations to a finite number of loop corrections, neglecting derivative terms, and adherence to CP symmetry yields a polynomial form for non-minimal coupling,  with even power of $\vp$, making the form in~\cref{eq:form of OmegaSq} the simplest and commonly adopted choice~\cite{AlHallak:2022gbv,Hertzberg:2010dc}.
}
\eq{\label{eq:form of OmegaSq}
\Omega^2(\vp) = 1+ \xi \frac{\vp^2}{\mpl^2}\,,
}
where both $\Omega^2$ and $\xi$ are dimensionless. 
In Einstein frame, the gravity sector and $\vp$ are not coupled. 
The metric in Einstein frame, denoted by $g^E_{\mu\nu}$, can be obtained as 
\begin{align}\label{eq:conformal}
g^E_{\mu\nu} = \Omega^{2} \, g^{JF}_{\mu\nu}\,.
\end{align}
In this article, Greek indices in both subscript and superscript range from $0$ to $3$. Additionally, we employ Einstein's summation convention for repeated indices. 
 In Einstein frame, the potential for inflaton is
\eq{\label{Eq:pot-Jordan_to_Einstein}
V^E(\varphi)= \frac{V^{JF}\lt(\vp\rt)}{\Omega^4 \lt(\vp\rt)}\,.
}
Furthermore, to make the kinetic term of inflaton canonical in Einstein frame, we need to redefine the inflaton as $\phi$. The relation between $\phi$ and $\vp$ is given by~\cite{Ghoshal:2023jvf}
\eq{\label{Eq:conversion-inflaton-EToJ}
\dv{\phi}{\varphi}=\sqrt{\Pi(\vp)}\,, 
}
where~\cite{Shaposhnikov:2020fdv,Racioppi:2019jsp,Park:2008hz}
\eq{
\Pi(\vp)= \frac{1}{\Omega^2} + \kappa \frac32 \lt( \frac{\mpl}{\Omega^2} \pdv{\qty(\Omega^2)}{\phi} \rt)^2
\,, 
\qquad \kappa=\begin{cases}
     1 \,\, \text{(metric formalism)}\,,\\
    0 \,\, \text{(Palatini formalism)}\,.
\end{cases}
}
 In metric formalism space-time metric, and other fields (for example, $\vp$ in~\cref{Eq:Jordan frame action}) are the independent dynamical variables, while in Palatini formalism $\Gamma_{\alpha \beta}^\lambda$ is also independent dynamical variable in addition to $\vp$ and space-time metric~\cite{Bauer:2008zj}. Moreover, in Palatini formalism Riemann tensor remains invariant under the transformation of~\cref{eq:conformal}~\cite{Bauer:2008zj}.  
 For the form of $\Omega^2$ mentioned in~\cref{eq:form of OmegaSq}, we have
\begin{empheq}[
  left=
    {
\dv{\phi}{\varphi}=
} 
    \empheqlbrace
]{align}
     &\sqrt{\frac{1+\lt(1+ 6\xi  \rt) \xi \vp^2/\mpl^2}{\lt(1+ \xi \vp^2/\mpl^2 \rt)^2} } \,\, &\text{(metric formalism)}\,, \label{Eq:conversion-inflaton-EToJ-metric}\\
    &\frac{1}{\sqrt{1+ \xi \vp^2/\mpl^2}} \,\, \,\;&\text{(Palatini formalism)}\,. 
    \label{Eq:conversion-inflaton-EToJ-Palatini}
\end{empheq}
Now, integrating~\cref{Eq:conversion-inflaton-EToJ-metric}, 
we get~\cite{Rasanen:2017ivk,Garcia-Bellido:2008ycs}
\eq{\label{Eq:Jordan-to-Einstein-field-conversion}
\sqrt{\xi}\frac{\phi}{\mpl} =
\sqrt{6 \kappa  \xi +1} \sinh ^{-1}\left(u \sqrt{  6 \kappa  \xi +1}\right)
-\sqrt{6 \kappa  \xi } \tanh ^{-1}\left(\frac{\sqrt{6 \kappa  \xi}  u}{\sqrt{u^2 (6 \kappa  \xi +1)+1}}\right)\,,
}
where $u=\sqrt{\xi}\frac{\vp}{\mpl}$. 
In metric formalism, under small field limit (i.e. $u\ll 1$ such that $\qty(1+u^2) \approx 1 $ and $\qty(1+u^2 +6 \xi u^2 )\approx 1$)~\cite{Bezrukov:2009db}, 
and under $\sqrt{\xi}u\gg 1$ we obtain~\cite{Bezrukov:2007ep}
\eq{\label{eq:metric-Einstein-approx-inflaton}
\vp \simeq \begin{cases}
    \phi \, \qquad &(\text{for }\sqrt{\xi}\vp\ll\mpl) \,,\\
    \frac{\mpl}{\sqrt{\xi}} \exp(\frac{1}{\sqrt{6}} \frac{\phi}{\mpl})\, \qquad &(\text{for } \xi \vp\gg \mpl)\,.
\end{cases}
}
However, in Palatini formalism, we can get exact relation between $\vp$ and $\phi$ from~\cref{Eq:conversion-inflaton-EToJ-Palatini} as follows  
\eq{\label{eq:Palatini-Einstein-exact-inflaton}
\vp=\frac{\mpl}{\sqrt{\xi}} \sinh(\sqrt{\xi}\frac{\phi}{\mpl})\,.
}
The relation between $\varphi$ and $\phi$ in~\cref{eq:metric-Einstein-approx-inflaton} is approximated, whereas the relation in~\cref{eq:Palatini-Einstein-exact-inflaton} is exact.
%
%

Moving forward, we assume that the interaction of inflaton with fermionic field $\chi$, which is singlet under \sm~gauge transformations, and with \sm~Higgs field $H$ during post-inflationary reheating era are defined in Einstein frame, and the interaction Lagrangian can be written as~\cite{Ghoshal:2022jeo,Ghoshal:2022aoh,Ghoshal:2023jvf,Ghoshal:2023noe}
\begin{align}\label{Eq:reheating lagrangian}
\mathcal{L}_{\rh} = - \yc \phi \cc - \lO \phi \HH 
- \lT \phi^2 \HH 
+ \text{h.c.}+\cdots  \,.
\end{align}
Among the three couplings $\yc,\lO,\lT$, only $\lO$ has mass dimension. Moreover, in~\cref{Eq:reheating lagrangian}, ellipses denote scattering of $\chi$ by SM particles or inflaton. However, we assume those interactions are not strong enough to keep $\chi$-particles in local-thermal equilibrium with the \sm~relativistic plasma of the universe. 
From~\Ccite{Bernal:2021qrl}, and also from our previous studies~\cite{Ghoshal:2023jvf,Ghoshal:2023jhh,Ghoshal:2022jeo}, it has been found that these scattering channels are not effective in producing enough $\chi$ particles such that total \cdm~density is satisfied, unless $\chi$ particles are highly massive (with a mass $\mc \sim {\cal O}\qty(10^{10}\GeV)$ or larger). As a result, our current focus in this article is solely on the production of 
$\chi$  particles through the decay of the inflaton. For the sake of completeness, the Lagrangian density of $\chi$ and $H$ are given below
\eq{
{\cal L}_\chi^E &= i \bar{\chi}\gamma^\mu {\partial_\mu} \chi - m_\chi \cc \,, \label{Eq:Lagrangian-for-chi}\\
{\cal L}_H^E &= \lt(\partial H\rt)^2 +\mH^2 \HH  - \lH \lt( \HH \rt)^2 \,,\label{Eq:Lagrangian-for-H}
}
where $i=\sqrt{-1}$, $\gamma^\mu$ are the four gamma matrices and $\bar{\chi}=\chi^\dagger\gamma^0$, and $\lH>0$. In~\cref{sec:reheating} of this article, we investigate whether $\chi$, vector-like fermionic stable particles produced through the decay of inflaton, contribute $100\%$ of the total \cdm~density of the present universe. In contrast, in~\cref{sec:lepto}, we explore the possibility of $\chi$ as a sterile yet unstable neutrino,  generating lepton asymmetry immediately after its production.


\section{Quartic potential}
\label{sec:Quartic slow roll}
In this article, we consider potential of inflaton in Jordan frame as~\cite{Gialamas:2023flv,Markkanen:2017tun,Takahashi:2018brt,Racioppi:2021jai}
\eq{\label{eq:Quartic potential in Jordan}
V^{JF}(\vp)=\Lambda \, \vp^4\,,
}
where $\Lambda$ is dimensionless. If we consider quantum loop correction arising due to the interaction of inflation with a fermionic and a scalar field, the potential for inflaton in Jordan frame is best described by Coleman-Weinberg potential (for example, see Ref.~\cite{Kannike:2016jfs,Racioppi:2021ynx,Okada:2015lia}).
However, this form of potential for slow roll inflation is in tension with the best-fit $n_s-r$ contour obtained from \Planck2018+\BICEP3~combined data~\cite{Racioppi:2017spw}, unless the renormalization scale is super Planckian~\cite{Ghoshal:2023jhh}. 
Since our focus in this work is to examine possible (in)consistencies, if any, 
in the metric and Palatini formalism, we restrict ourselves to considering only the simple quartic form of the inflationary potential.

The potential of inflation in Einstein frame can be written as
\begin{align}\label{eq:pot-Einstein}
    V^E(\varphi)=\frac{\Lambda \, \varphi^4}{\Omega^4(\varphi)}\,.
\end{align}

In metric formalism and for $\xi \vp \gg \mpl$, the potential for inflaton mentioned in~\cref{eq:pot-Einstein} is given by~\cite{Takahashi:2018brt} 
\eq{\label{eq:metric-pot-approx-large}
V^E_{(m)}(\phi)\simeq \frac{\Lambda \, \mpl^4}{\xi^2} \left(1-e^{-\sqrt{\frac{2}{3}} \frac{\phi }{\mpl}}\right)^2\,,
}
and under small field limit (from~\cref{eq:metric-pot-approx-large})~\cite{Takahashi:2018brt,Garcia-Bellido:2008ycs}
\eq{\label{eq:pot-metric-small}
V^E_{(m)}(\phi)\approx \frac{2}{3}\frac{\Lambda }{\xi^2}\mpl^2\, \phi^2\,.
} 
In Palatini formalism, the potential of inflaton in Einstein frame~\cite{Gialamas:2023flv,Takahashi:2018brt} 
\eq{\label{eq:potential-in Einstein-frame}
V^E_{(P)}(\phi)= \mpl^4 \frac{\Lambda}{\xi^2} \qty(\tanh(\sqrt{\xi} \frac{\phi}{\mpl}))^4\,.
}
Under small field approximation (assuming $\sqrt{\xi}\phi/\mpl <\pi/2$)~\cite{Takahashi:2018brt}
\eq{\label{eq:pot-Palatini-small}
V_{(P)}^E(\phi)\approx \Lambda \phi^4\,.
}

From~\cref{eq:pot-metric-small,eq:pot-Palatini-small} we see that during reheating when inflaton oscillates about the minimum, the potential is quadratic in metric formalism and quartic in Palatini formalism~\cite{Takahashi:2018brt}. 

Under slow roll approximation, the observables of the inflationary epoch, namely scalar spectral index ($n_s$), tensor to scalar ratio ($r$), and amplitude of comoving curvature power spectrum ($A_s$) are defined in Einstein  frame as follows~\cite{Baumann:2009ds}
\begin{align}
    n_s(\phi_*) \seq 1-6\, \epsilon_V^E(\phi_*) + 2 \, \eta_V^E(\phi_*)\,, \quad & r(\varphi_*) \approx 16 \, \epsilon_V^E(\phi_*)\,, &V^E (\phi_*)\seq \frac32 \pi^2 \, r(\phi_*) \, A_s\, \mpl^4\,.
\end{align}
Here, $\phi_*$ is the value of inflaton corresponding to the pivot scale of \cmb~observations, whereas $\epsilon_V^E(\phi)$, and $\eta_V^E(\phi)$ are the potential slow roll parameters defined as~\cite{Lyth:2009imm,Okada:2015lia} $\epsilon_V^E(\phi)=  \qty({\mpl^2}/{2})\, \qty({\dd \ln [V^E]}/{\dd \phi})^2$ and $\eta_V^E(\phi)=\qty(\mpl^2/V^E) \qty( {\dd^2 V^E}/{\dd \phi^2})$. During slow roll inflationary epoch $\epsilon_V^E,\eta_V^E\ll 1$.  Current bounds on $n_s$, $r$, and $A_s$ are mentioned in~\cref{Table:PlanckData}. The duration of the inflationary epoch is expressed in terms of the number of e-folds ($\ncmb$) which is defined as~\cite{Baumann:2009ds}
\begin{align}
    \ncmb=\mpl^{-1}\int^{\phi_*}_{\phi_{\rm end}} \qty(2\,   \epsilon_V^E)^{-1/2} \, \dd \phi\,,
\end{align}
where $\phi_{\rm end}$ is the value of the inflaton at which the slow roll phase ends, i.e., when the kinetic energy of the inflaton becomes $\sim V^E(\phi)$. In other words, it happens when any of $\epsilon_V^E$ or $\eta_V^E$ becomes $\sim 1$~\cite{Ghoshal:2022jeo}. In this article, we consider $50\leq \ncmb \leq 60$.  
\begin{table}[H]
\begin{center}
\caption{ \centering\it Constraints on $n_s,r$ and $A_s$ (T and E for temperature and E-mode polarization of \cmb).} \label{Table:PlanckData}
\begin{tabular}{ |c|| c| c||c| }
\hline
$\ln(10^{10} A_s)$ & $
3.044\pm 0.014$ & $68\%$, TT,TE,EE+lowE+lensing+BAO & 
\cite{Planck:2018vyg,ParticleDataGroup:2022pth}  \\
 \hline
 $n_s$ & $0.9647\pm 0.0043$ & $68\%$, TT,TE,EE+lowE+lensing+BAO & 
 \cite{Planck:2018vyg} \\ 
 \hline 
 $r$ & $0.014^{+0.010}_{-0.011}\, \text{and}$ &  $ 95 \%  \,, \text{BK18, \textsc{Bicep}3, \textit{Keck Array}~2020,}$& \cite{Planck:2018vyg, BICEPKeck:2022mhb,BICEP:2021xfz}\\
   & $ <0.036 $ & and \textsl{WMAP} and \textit{Planck}~CMB polarization & (see also \cite{Campeti:2022vom}) \\
 \hline 
\end{tabular}
\end{center}
\end{table}%
%
%
%
%

Using the relation between $\phi$ and $\varphi$ from~\cref{eq:Palatini-Einstein-exact-inflaton} (for Palatini) and {exact expression from~\cref{Eq:Jordan-to-Einstein-field-conversion} (for metric), while varying $\xi$, we obtain benchmark values for slow roll inflation for both the metric and Palatini formalisms}. These benchmark values satisfy the bounds presented in~\cref{Table:PlanckData} and are obtained for $\ncmb\sim 50$ and $60$. These benchmarks are listed in~\cref{table:quartic-Benchmark-Palatini,table:quartic-Benchmark-metric} ($\varphi_*$ and $\varphi_{\rm end}$ in \jframe~correspond to the value of $\phi_*$ and $\phi_{\rm end}$ in \eframe). Here, we are mainly interested in  the regime where $\xi\gg 1$, as in this case, the predictions between the metric and Palatini formalisms exhibit significant differences~\cite{Cheong:2021kyc}. This will help us make the kind of comparison we intend to do between the two formalisms.
In either of the metric or Palatini formalism, 
~\cref{table:quartic-Benchmark-metric,table:quartic-Benchmark-Palatini} indicate that for a constant value of $\xi$, the value of $\varphi_{\rm end}$ (or $\phi_{\rm end}$) remains fixed.


\begin{table}[H]
\centering
\caption{Benchmark values for slow roll inflation in metric formalism.}
\label{table:quartic-Benchmark-metric}
\begin{tabular}{|c || c | c| c|c|c|c|c|} 
 \hline
 Benchmark & $\xi$ & $\vp_{\rm end}/\mpl$ & $\varphi_*/\mpl$ & $\ncmb$ & $n_s$ & $r\times 10^3$ & $\Lambda$ \\ [0.5ex]
 \hline\hline
(m)BM1 & \multirow{2}{*}{$0.5$} & \multirow{2}{*}{$1.3287$} & $11.2724$ & $60$ & $0.9677$ & $3.9483$ & $3.1644\times 10^{-11}$ \\ 
\cline{1-1}\cline{4-8}
(m)BM2  &  &  & $10.3349$ & $50$ & $0.9614$ & $ 5.5837$ & $4.5015\times 10^{-11}$\\
  \hline 
\mmoon  & \multirow{2}{*}{$10$}  & \multirow{2}{*}{$0.3372$}  & $2.8886$  & $60$ & $0.9678$  & $ 3.0132$ & $9.5886\times 10^{-9}$\\
\cline{1-1}\cline{4-8}
(m)BM4  &  &  & $2.6488$ & $50$ & $0.9616$ & $4.2620$& $1.3624\times 10^{-8}$ \\
  \hline 
(m)BM5  & \multirow{2}{*}{$10^2$} & \multirow{2}{*}{$0.1074$}  & $ 0.9203$ & $60$ & $0.9678$& $2.9688$ & $9.4441\times 10^{-7}$ \\
\cline{1-1}\cline{4-8}
(m)BM6  &  &  & $0.8439$ & $50$ & $0.9616$ & $4.1993$ & $1.3417\times 10^{-6}$\\
  \hline 
\msun  & \multirow{2}{*}{$10^4$} & \multirow{2}{*}{$0.0107$}  & $ 0.0921$ & $60$ & $0.9678$& $2.9639$ & $9.4283\times 10^{-3}$ \\
\cline{1-1}\cline{4-8}
(m)BM8  &  &  & $0.0845$ & $50$ & $0.9616$ & $4.1924$ & $1.3395\times 10^{-2}$\\
  \hline 
\end{tabular}
\end{table}

\begin{table}[H]
\centering
\caption{Benchmark values for slow roll inflation in Palatini formalism.}
\label{table:quartic-Benchmark-Palatini}
\begin{tabular}{|c || c | c| c|c|c|c|c|} 
 \hline 
 Benchmark & $\xi$ & $\vp_{\rm end}/\mpl$ & $\varphi_*/\mpl$ & $\ncmb$ & $n_s$ & $r\times 10^3$ & $\Lambda$ \\ [0.5ex]
 \hline\hline
 (P)BM1 & \multirow{2}{*}{$0.5$} & \multirow{2}{*}{$1.7672$} & $21.9801$  & $60$ & $0.9668$ & $1.0923$ & $8.5554\times 10^{-12}$\\ 
 \cline{1-1}\cline{4-8}
(P)BM2  &  &  & $20.0779$ & $50$ & $0.9602$ & $1.5675$ & $1.2298\times 10^{-11}$\\
  \hline 
\Pmoon  & \multirow{2}{*}{$10$}  &  \multirow{2}{*}{$0.9197$} & $21.9282$ & $60$ & $0.9667$ & $5.5349\times 10^{-2}$ & $1.7206\times 10^{-10}$ \\
 \cline{1-1}\cline{4-8}
(P)BM4  & &   & $20.0211$ & $50$ & $0.9601$ & $7.9643\times 10^{-2}$ & $2.4760\times 10^{-10}$\\
  \hline 
(P)BM5  & \multirow{2}{*}{$100$} & \multirow{2}{*}{$0.5271$} & $21.9152$ & $60$ & $0.9667$ & $5.5490\times 10^{-3}$ & $1.7243\times 10^{-9}$\\
 \cline{1-1}\cline{4-8}
  (P)BM6  &  &   & $20.0069$ & $50$ & $0.9600$ & $7.9887 \times 10^{-3}$ & $2.4825\times 10^{-9}$\\
  \hline 
  \Psun  & \multirow{2}{*}{$10^4$} & \multirow{2}{*}{$0.1680$} & $21.9095$ & $60$ & $0.9667$ & $5.5549\times 10^{-5}$ & $1.7261\times 10^{-7}$\\
 \cline{1-1}\cline{4-8}
  (P)BM8  &  &  & $20.0007$ & $50$ & $0.9600$ & $7.9989 \times 10^{-5}$ & $2.4855\times 10^{-7}$\\
  \hline 
\end{tabular}
\end{table}

\begin{figure}
    \centering
    \includegraphics[width=0.5\linewidth]{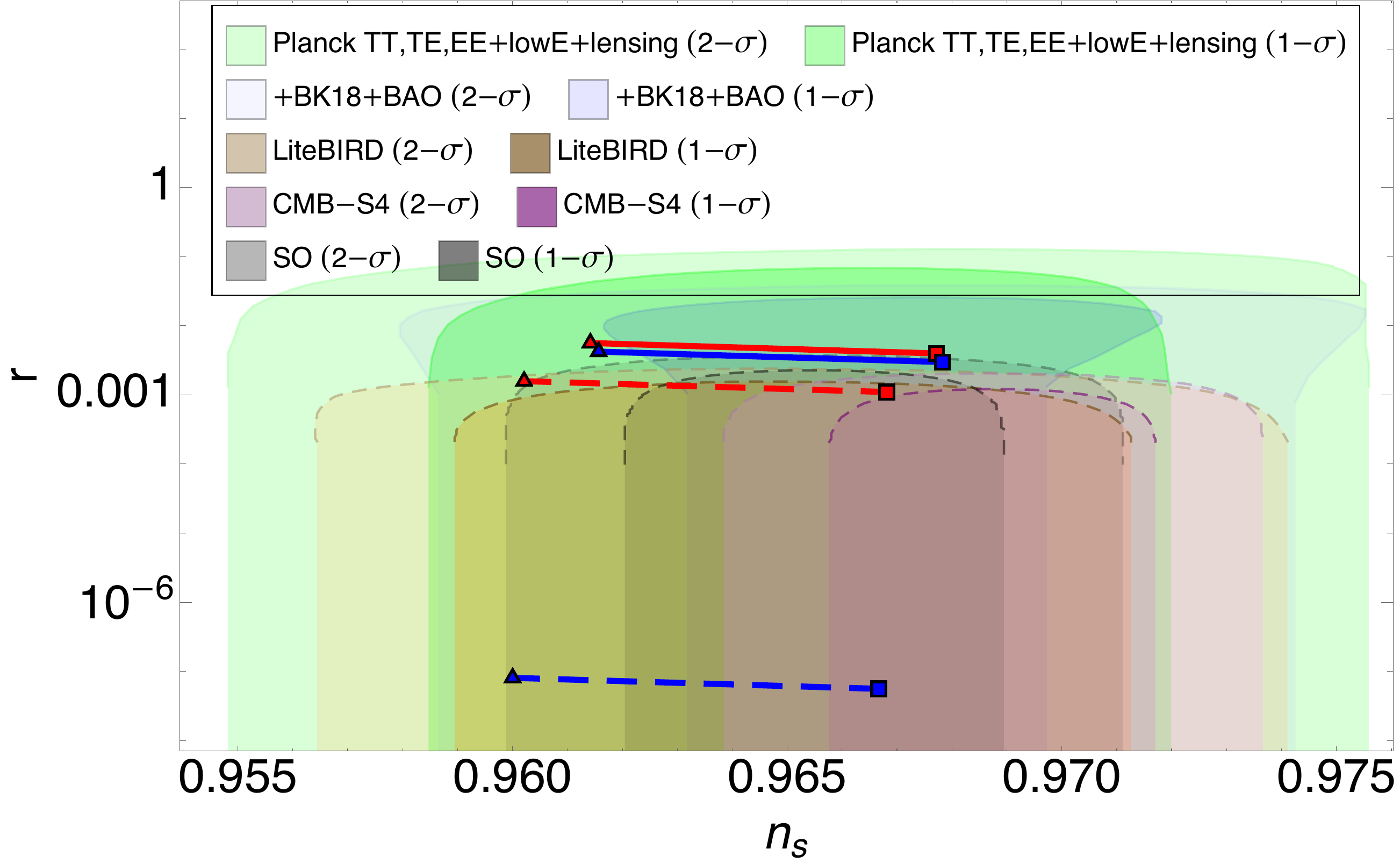}
    \caption{{\it
    \raggedright
    $(n_s-r)$ predictions for $\xi=0.5$ and $\xi=10^4$ in both metric and Palatini formalism and for $50\lsim \ncmb\lsim 60$ along with bounds on  $(n_s-r)$ plane from \cmb~observations (\Planck2018, \BICEP3, \KeckArray) and prospective future reach from upcoming \cmb~experiments (\LB,\cmbsfour,SO~\cite{LiteBIRD:2020khw,CMB-S4:2016ple,SimonsObservatory:2018koc}). The solid straight lines correspond to $(n_s,r)$ predictions in metric formalism - red color line for $\xi=0.5$ and blue colored line for $\xi=10^4$. The dashed straight lines correspond to $(n_s,r)$ predictions in Palatini formalism - red colored line for $\xi=0.5$ and blue colored line for $\xi=10^4$. Along the lines, $\ncmb$ is varied from $50$ to $60$ from left to right.
    }
    }
    \label{fig:ns-r-contour}
\end{figure}

\begin{figure}[H]
    \centering
    \includegraphics[width=0.45\linewidth]{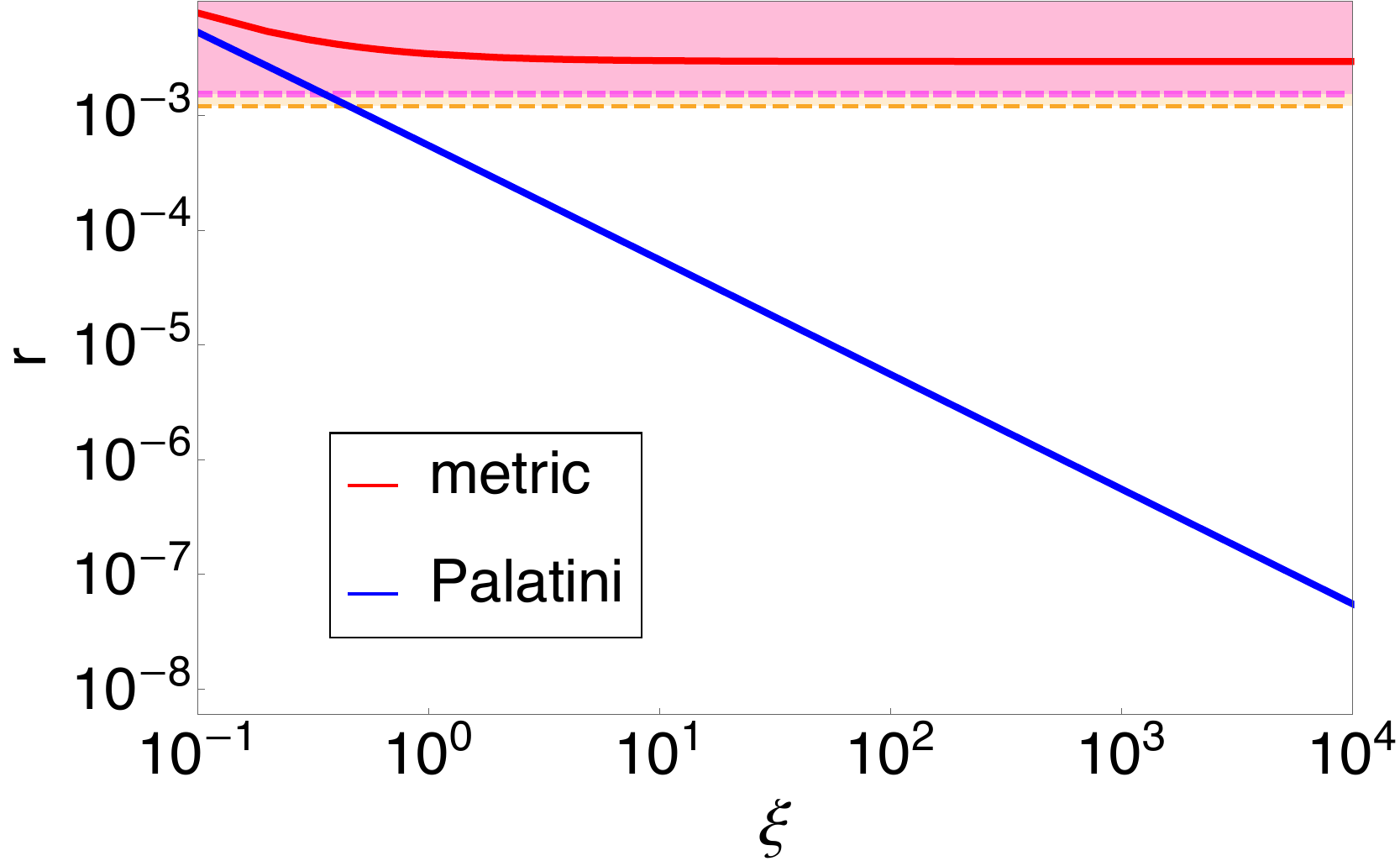}
\includegraphics[width=0.45\linewidth]{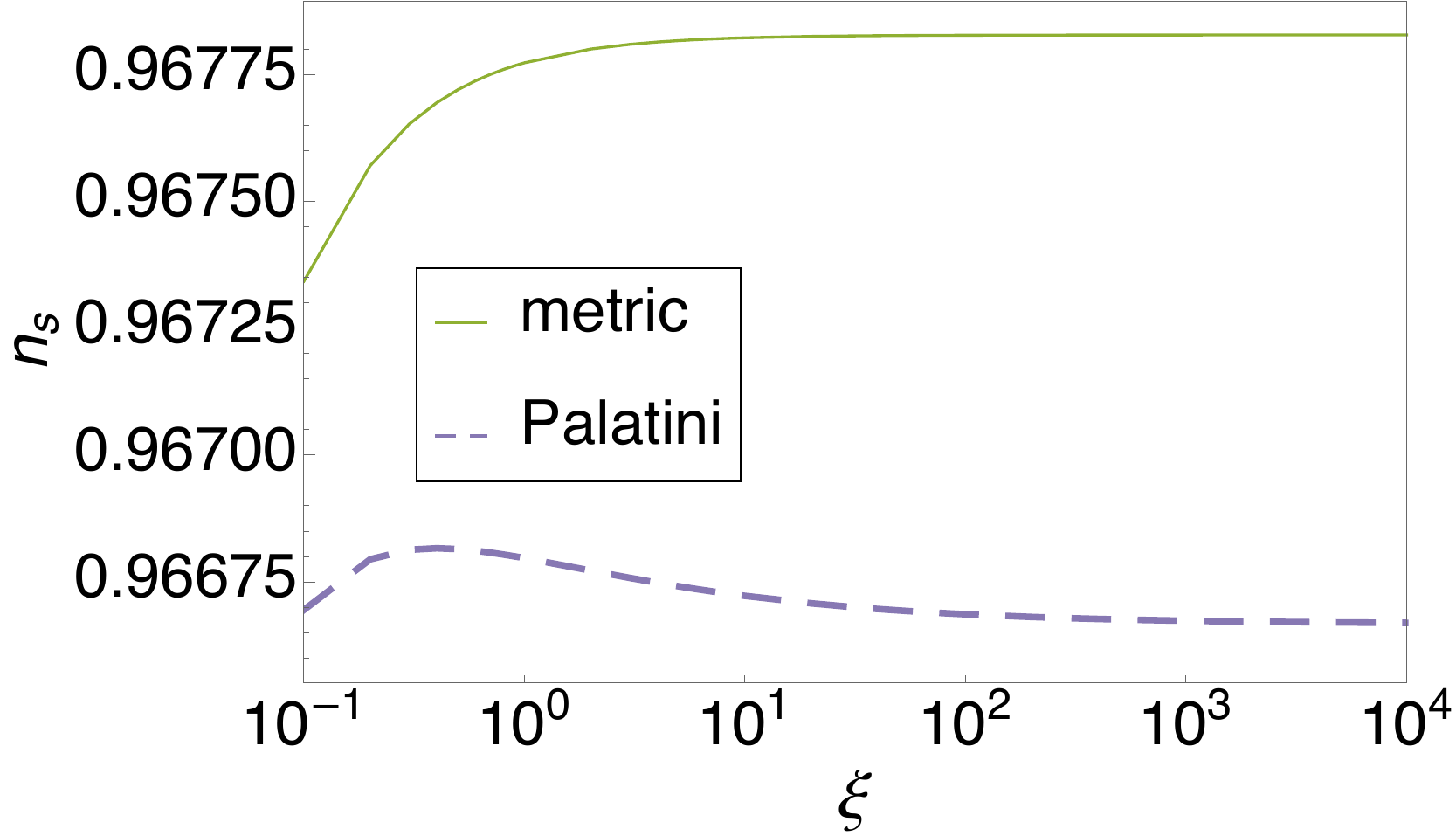}
        \caption{\it
    \raggedright
{\bf Left-pannel:} Variation of predicted value of $r$ against $\xi$ for $\ncmb=60$ for metric and Palatini formalisms. The pink-colored and Yellowish orange-colored regions represent the values of $r$ which can be validated at $1-\sigma$ CL by the upcoming CMB experiments \LB, and \cmbsfour, respectively. {\bf Right-pannel:} Variation of the predicted value of $n_s$ against $\xi$ for $\ncmb=60$ for metric and Palatini formalisms. 
}
    \label{fig:ns-r-versus-Xi}
\end{figure}

\cref{fig:ns-r-contour} displays the $n_s-r$ predictions for slow roll inflation corresponding to the potential of the inflaton mentioned in~\cref{eq:pot-Einstein}, in both metric and Palatini formalisms. The solid straight lines are for predictions in metric formalism:  red colored line for $\xi=0.5$, while blue colored line for $\xi=10^4$. The dashed straight lines are for predictions in Palatini formalism:  red colored line for $\xi=0.5$, while blue colored line for $\xi=10^4$. Along these lines, $\ncmb$ varies between $50$ and $60$, marked with a triangular symbol at $\ncmb \simeq 50$ and a square symbol at $\ncmb \simeq 60$. \cref{fig:ns-r-contour} also shows bounds on the $n_s-r$  plane from observed \cmb~data from \Planck2018 and \Planck2018+\BICEP3+\KeckArray2015 combined analysis, along with future prospective reaches for upcoming \cmb~experiments (with dashed curved lines as perimeter) e.g. \cmbsfour, \LB, Simons Observatory (SO), shown in the background. This figure shows that 
while the predicted values of $(n_s,r)$ in both the metric and Palatini formalisms, for both $\ncmb = 50$ and $\ncmb = 60$, fall within the $1-\sigma$ best-fit contour of \Planck2018 data, those predicted values for $\ncmb$ around $50$ ($\ncmb\gsim 50$) are excluded at the $1-\sigma$ CL by the combined analysis of \Planck+\BICEP+\KeckArray~data. 
Furthermore, \LB,\cmbsfour, and SO can verify all $(n_s, r)$ values, particularly those predicted within the metric formalism. For example, predicted values of $(n_s,r)$ in metric formalism for $\ncmb\sim 60$ can be tested by upcoming SO at $1-\sigma$ CL. 
For the predicted values of $(n_s,r)$ in Palatini formalism around $\ncmb\sim 50$ for $\xi=0.5$ can be tested in the future by \LB~at $1-\sigma$ CL or by SO at $2-\sigma$ CL, and around $\ncmb\sim 60$ for $\xi=0.5$ can be verified by upcoming \cmbsfour~at $1-\sigma$ CL. The predicted values of $(n_s,r)$ in Palatini formalism around $\ncmb\sim 50$ for $\xi=10^4$ can be validated in the future by forthcoming SO~at $1-\sigma$ CL.
%
From~\cref{table:quartic-Benchmark-metric,table:quartic-Benchmark-Palatini} and~\cref{fig:ns-r-contour}, we see that predicted values $r$, at a fixed value of $\xi$, exhibit a decrease as the value of $\ncmb$ increases. This is due to $\ln\lt(r\rt) \propto - \ln\lt(\ncmb\rt)$ observed in both formalisms~\cite{Cheong:2021kyc}. 
Furthermore,~\cref{fig:ns-r-versus-Xi}  depicts variation of $n_s$ and $r$ with $\xi$ for $\ncmb\sim60$. Comparatively, $r$ exhibits  more significant variation with $\xi$ than $n_s$, specifically, in Palatini formalism. This can be explained as follows: when $\xi\gg 1$, at leading order approximation $r\sim \xi^0$ in metric formalism, while $r\sim \xi^{-1}$ in Palatini formalism~\cite{Cheong:2021kyc}. Furthermore, in the left panel, the areas shaded in pink and yellowish-orange depict the values of $r$ which can be validated at $1-\sigma$ CL by the forthcoming CMB experiments LiteBIRD and CMB-S4, respectively. Therefore, this future \cmb~observation can refute or validate at $1-\sigma$ CL the analysis of slow roll inflation in metric formalism for the plain quartic potential of inflaton and with non-minimal coupling with Ricci scalar.

\section{Stability}
\label{sec:stability}
In this section, we explore the permissible upper limit of $\yc$ and $\lO$ such that the radiative loop correction arising due to coupling of inflaton with $\chi$ and $H$ does not distort the flatness of the potential $V^E_{(m)}$ or  $V^E_{(P)}$. 
Actually, we focus on reheating via $\lambda_{12}$, assuming a very small permissible value for $\lambda_{22}$.
Here, we use exact expression of $V^E_{(m)}$ rather than relying on an approximate expression, such as the expression given in~\cref{eq:metric-pot-approx-large}. In this section, we use $V^E_{\rm tree}$ to denote  $V^E_{(m)}$ or  $V^E_{(P)}$ and call them as tree level potential. The Coleman–Weinberg (CW) correction to the inflaton potential is~\cite{Coleman:1973jx}
\begin{align}\label{eq:loop correction}
V_{\lop}
=\sum_{j
} \frac{n_j}{64\pi^2} (-1)^{2s_j}\widetilde{m}_j^4
\left[ \ln\left( \frac{\widetilde{m}_j^2 
}{\mu^2} \right) - c_j \right]  \,.
\end{align}
Here, inflaton dependent mass for $j\equiv \chi$ and for $j=H$ are $\widetilde{m}_{\chi}^2 (\phi) = \left( m_\chi + y_\chi \phi \right)^2 $ and $\widetilde{m}_{H}^2 (\phi) = m_H^2 + \lambda_{12} \phi$, respectively. Furthermore, $c_j=\frac{3}{2}$; $n_{H,\chi}=4$; $s_H =0$, $s_\chi=1/2$; and, we consider two values of the renormalization scale: $\mu=\phi_{*}$ and $\mu=\phi_{\rm end}$. Now, the first and 
second derivatives of the Coleman–Weinberg term of~\cref{eq:loop correction} for $\chi$ and $H$ with respect to $\phi$ are~\cite{Drees:2021wgd}
\begin{align}
	 & V_{\lop}^\prime = \sum\limits_{j 
	 }\frac{n_j}{32 \pi^2}  (-1)^{2 s_j} 
	\widetilde{m}_j^2 \, \lt(\widetilde{m}_j^{2}\rt)^{\prime}
	\left[ \ln \left(\frac{\widetilde{m}_j^2}{\mu^2} \right)
	- 1 \right]\,,  \label{Eq:first derivative test}\\
	& V_{\lop}^{\prime\prime} =  \sum\limits_{j
	} \frac{n_j}{32 \pi^2} (-1)^{2 s_j} 
	\left\{ \left[ \left(\left(\widetilde{m}_j^{2}\right)^{\prime} \right)^2
	+ \widetilde{m}_j^2 \left(\widetilde{m}_j^{2}\right)^{\prime\prime} \right]
	\ln \left(\frac{\widetilde{m}_j^2}{\mu^2} \right)
	- \widetilde{m}_j^2 \left(\widetilde{m}_j^{2}\right)^{\prime\prime}  \right\}\,.   \label{Eq:second derivative test}
\end{align}
%
The upper permissible values of $\yc$ and $\lO$ can be obtained if both of the following conditions are simultaneously satisfied  (with $\mc\approx \mH\approx 0$) 
\eq{
 \lt| V_{{\lop},j}^\prime (\phi=\mu) \rt|<V_{\rm tree}^{\prime} (\phi=\mu) \,,
 &&
  \lt|V_{{\lop},j}^{\prime \prime} (\phi=\mu)\rt| <V_{\rm tree}^{\prime\prime} (\phi=\mu)\,.
}
From this point forward, we work with two benchmark values  corresponding to $\xi=10$ and $\xi=10^4$ to demonstrate the dependence on the results on the coupling parameter. We  select these benchmark values with $\xi\gg 1$ to distinctly differentiate between the metric and Palatini formalisms, as discussed earlier. Furthermore, we select benchmarks corresponding to $\ncmb = 60$, as predicted $(ns, r)$ values fall within the $1 - \sigma$ contour of \Planck+\BICEP~combined analysis on the $n_s-r$ plane. The upper limit of $\yc$ and $\lO$ from stability analysis are mentioned in~\cref{Table:Quartic_metric_stability} for metric formalism (Benchmark: \mmoon~and \msun) and in~\cref{Table:Quartic_Palatini_stability} for Palatini formalism (Benchmark: \Pmoon~and \Psun).

\begin{table}[H]
\begin{center}
\caption{Allowed upper limit of $\yc$ and $\lO$ for the benchmark values \mmoon~and \msun~from~\cref{table:quartic-Benchmark-metric}.}
\label{Table:Quartic_metric_stability}
\begin{tabular}{|c|| c| c| c|c|} 
 \hline
{\it Benchmark}  &\multicolumn{2}{|c|}{stability for $\yc$} & \multicolumn{2}{|c|}{stability for $\lO$}\\ [0.5ex] 
  \cline{2-5} 
   & about $\mu=\phi_*$ & about $\mu=\phi_{\rm end}$ & about $\mu=\phi_*$ & about $\mu=\phi_{\rm end}$ \\ [0.5ex] 
 \hline\hline
\mmoon & $\yc<4.0723\times 10^{-4}$ & $\yc<1.4026
 \times 10^{-3}$ & $\lO/\mpl<1.3157\times 10^{-6}$ & $\lO/\mpl<1.7633\times 10^{-5}$ \\ 
 \hline
\msun & $\yc<4.8013\times 10^{-4}$ & $\yc<2.0114\times 10^{-3}$ & $\lO/\mpl<1.8197\times 10^{-6}$ & $\lO/\mpl<3.8608\times 10^{-5}$ \\
 \hline
\end{tabular}
\end{center}
\end{table}

\begin{table}[H]
\begin{center}
\caption{Allowed upper limit of $\yc$ and $\lO$ for the benchmark values \Pmoon~and \Psun~ from~\cref{table:quartic-Benchmark-Palatini}.}
\label{Table:Quartic_Palatini_stability}
\begin{tabular}{|c|| c| c| c|c|} 
 \hline
{\it Benchmark}  &\multicolumn{2}{|c|}{stability for $\yc$} & \multicolumn{2}{|c|}{stability for $\lO$}\\ [0.5ex] 
  \cline{2-5} 
   & about $\mu=\phi_*$ & about $\mu=\phi_{\rm end}$ & about $\mu=\phi_*$ & about $\mu=\phi_{\rm end}$ \\ [0.5ex] 
 \hline\hline
\Pmoon & $\yc<2.8987\times 10^{-4}$ & $\yc<1.2861
 \times 10^{-3}$ & $\lO/\mpl<1.0050\times 10^{-7}$ & $\lO/\mpl<7.8871\times 10^{-6}$ \\ 
 \hline
\Psun & $\yc<2.0864\times 10^{-4}$ & $\yc<2.4894\times 10^{-3}$ & $\lO/\mpl<1.3483\times 10^{-8}$ & $\lO/\mpl<1.7057\times 10^{-6}$ \\
 \hline
\end{tabular}
\end{center}
\end{table}

From~\cref{Table:Quartic_metric_stability,Table:Quartic_Palatini_stability} we infer the following upper limits of $\lO$ and $\yc$ (i.e. ${\lO}_{,{\rm max}}$, and ${\yc}_{,{\rm max}}$) that respect the stability criteria:
for \mmoon, $\yc< 4.0723\times 10^{-4}$ and $\lO/\mpl< 1.3157 \times 10^{-6}$; for \msun, $\yc< 4.8013\times 10^{-4}$ and $\lO/\mpl< 1.8197\times 10^{-6}$; for \Pmoon, $\yc< 2.8987\times 10^{-4}$ and $\lO/\mpl< 1.0050\times 10^{-7}$; and for \Psun, $\yc< 2.0864\times 10^{-4}$ and $\lO/\mpl< 1.3483\times 10^{-8}$.

\section{Reheating and production of $\chi$ from inflaton decay}
\label{sec:reheating}
The inflationary epoch is succeeded by the reheating era, during which the adiabatic production of \sm~and possibly \bsm~particles from the oscillating inflaton takes place. Consequently, the universe transitions from  a cold to a hot radiation-dominated state as the inflaton decays. With the formation of radiation (relativistic \sm~particles), the temperature of the universe rises to its maximum ($\Tmax$) and  subsequently decreases due to Hubble expansion. Actually, in the early stages of reheating, the total reaction rate ($\Gamma_\phi$) of the inflaton is lower than the Hubble parameter ($\hubble$), with the latter being the primary factor causing a decrease in the energy density of the oscillating inflaton. As the universe continues to expand, at reheat temperature ($\Trh$), $\hubble$ becomes comparable to $\Gamma_\phi$, resulting in $\rho_\phi = \rho_\rad$, indicating equality in energy density between the oscillating inflaton ($\rho_\phi$) and radiation ($\rho_\rad$). Following this, the remaining energy density of inflaton quickly converts to \sm~and \bsm~particles, marking the beginning of radiation radiation-dominated universe. In this work, we investigate perturbative reheating, and we assume that the \sm~particles generated during this epoch rapidly achieve thermal equilibrium soon after their formation.

The equation of state parameter of the oscillating inflaton, $w_{\phi}$, can be defined as $ w_\phi=P_\phi/\rho_\phi$%
\footnote{{We expect that $w_\phi \in [-1,0]$ in Einstein frame in both metric and Palatini formalism during slow-roll inflationary phase, and $w_\phi \in [0,1]$ during reheating era ($w_\phi=1$ at the bottom of the potential at the location of the minimum of the potential of inflaton). However, exact value of $w_\phi$ is determined by the particle physics details of the reheating process (see~\Ccite{Cheong:2021kyc}). Nonetheless, claiming this in the Jordan frame is not straightforward, as the kinetic energy is not canonical here (see~\Ccite{Jarv:2024krk}).}}
\begin{align}\label{eq:w-k-relation}
    w_\re \simeq \frac{k-2}{k+2}\,.
\end{align}
Therefore, from~\cref{eq:pot-metric-small,eq:pot-Palatini-small}, we can infer that
\begin{align}
    w_\re\simeq \begin{cases}
        0 \, \quad &\text{(in metric formalism)}\,,\\
        \frac13 \, \quad &\text{(in Palatini formalism)}\,.
    \end{cases}
\end{align} 
Hence, in metric formalism, inflaton behaves as non-relativistic fluid, whereas in Palatini formalism inflaton behaves as relativistic fluid. 
The primary reason behind this is that in Palatini formalism, $V_{(P)}^E(\phi)\sim \phi^4$, making $\phi$ massless and causing its evolution as radiation, i.e. $\rho_\phi\sim \cs^{-4}$~\cite{Garcia:2020eof}.
From now on, we will use ``(in metric formalism)" and ``(in Palatini formalism)" to provide a side-by-side comparison of the results obtained using the condition that the potential of inflaton around the minimum is quadratic in metric formalism and quartic potential in Palatini formalism. Note that this difference in results might not be a generic feature for every model of inflation. In certain cases, the potential around the minimum could be quadratic in both the metric and Palatini formalisms. However, that we can find significant difference between the two formalisms for a class of inflationary models is a salient point of the present analysis, that needs to be explored further as we go along.

Following the discussion of~\cref{appendix:Boltzmann}, we define $\Trh$ in both metric and Palatini formalisms as
\begin{align}
    \Trh 
\simeq &\sqrt{\frac{2}{\pi}} \left(\frac{10}{\gs}\right)^{1/4} \sqrt{\mpl} \sqrt{\Gamma_\phi}\,,   \label{Eq:definition of reheating temperature}
\end{align}
where $\Gamma_\phi$ is the total decay width of inflaton. 
This particular choice in defining $\Trh$ ensures that $\rho_\phi(\Trh)=\rho_\rad(\Trh)$. However, $\Tmax$ depends on whether the potential of inflaton is quadratic or quartic during reheating. %
$\Tmax/\Trh$ for the quadratic and quartic potential of inflaton during reheating is derived in~\cref{appendix:Boltzmann}. Since, the potential about the minimum of the inflaton is quadratic in metric formalism, and quartic in Palatini formalism for our considered inflationary scenario, we obtain from~\cref{eq:TMAX-quadratic} 
and~\cref{eq:TMAX-quartic}~\cite{Bernal:2019mhf,Giudice:2000ex,Garcia:2020eof}
\begin{empheq}[
  left=
    {
\frac{\Tmax}{\Trh}\simeq
} 
    \empheqlbrace
]{align}
&\lt(\frac38\rt)^{2/5} \, \lt(\frac{\hubble_I}{\hubble(\Trh)}\rt)^{1/4}\,\quad &&\text{(in metric formalism)}\,,\\
&\frac{3^{1/4}}{2} \, \lt(\frac{\hubble_I}{\hubble(\Trh)}\rt)^{1/4}\,\quad &&\text{(in Palatini formalism)}\,.
\end{empheq}

Here, we assume the {\it effective number of relativistic degrees of freedom} $\gs\approx 106.75$. 
Furthermore, ${\cal H}_I$ denotes the value of Hubble parameter at the end of inflation
(i.e. when $\rho_{\rm rad}=0$~\cite{Giudice:2000ex}), which is given by
\be\label{Eq:HI-natural}
{\cal H}_I\simeq\sqrt{\frac{V^E\qty(\phi_{\rm end})}{3\mpl^2}}\,.
\ee
Furthermore, $\hubble(\Trh)$ is the value of Hubble parameter when $T=\Trh$, i.e. 
\begin{align}
    \hubble(\Trh)\simeq \frac{\pi}{3\mpl} \sqrt{\frac{\gs}{10}} \,\Trh^2\,.
\end{align}
Considering~\cref{Eq:reheating lagrangian}, $\Gamma_\phi$ depends on decay width of inflaton to \sm~Higgs particle ($h$), denoted as $\Gamma_{\phi\to {h h}}$, and \bsm~particle ($\chi$), denoted as $ \Gamma_{\phi \to \chi\chi}$. These are given by  
\eq{
&\G_\phi=\Gamma_{\phi\to {h h}}+ \Gamma_{\phi \to \chi\chi} \approx \Gamma_{\phi\to {h h}}\,,\label{Eq:decay-width-of-inflaton}\\
& \Gamma_{\phi\to {h h}}  \simeq \frac{\lambda_{12}^2}{8\pi\, m_{\phi}}\,, 
 \qquad 
 \Gamma_{\phi \to \chi\chi}  \simeq \frac{y_\chi^2\, m_{\phi}}{8\pi}\,.\label{eq:decay width}
}
The reason for using the approximate sign in~\cref{Eq:decay-width-of-inflaton} is to prevent the universe from being dominated by \dm~immediately after reheating era. In~\cref{eq:decay width} $m_\phi$ is the effective mass of inflaton in Einstein frame, which can be obtained for both metric and Palatini formalisms 
as
\eq{
m_\phi^2= \left.\frac{\td^2 V^E}{\td \phi^2} \right|_{\phi=\phi_{\rm min}} \,.
} 
The value of $m_\phi=0$ in both formalisms because the potential from~\cref{Eq:pot-Jordan_to_Einstein} has a minimum at $\phi=0$.
However, we can always add a bare mass term ${{\baremass}}^2\, \phi^2$ to the potential.  
To ensure that the slow roll inflationary scenario discussed in~\cref{sec:Quartic slow roll}, holds good even after the inclusion of this bare mass term,
an upper limit is imposed on the bare mass, and it has been discussed in~\cref{appendix:bare mass}. %
%
%
%
%
%
%
By using~\cref{eq:pot-metric-small,eq:appendix-bare-mass-upper-limit-metric,eq:appendix-bare-mass-upper-limit-Palatini}
we prefer to choose the value of bare mass as follows 
\begin{align}
&{\baremass}_{(m),\xi}=\frac{1}{\upzeta} \, \sqrt{\frac{2}{3}\frac{\Lambda }{\xi^2}\mpl^2}\,,\label{eq:baremass-metric}\\
   &{\baremass}_{{(P)},\xi} \simeq  \frac{1}{\upzeta}\, \sqrt{\Lambda} \lt(\frac{\sqrt{3}\, \mpl\, \lambda_{12}^2}{8\pi \,{\Lambda}} \rt)^{1/3}
   =\frac{1}{\upzeta} \,  \lt(\frac{3 \pi^2}{40 }\rt)^{1/4}  \, \qty(\gs \, \Lambda)^{1/4} \, \Trh\,, \label{eq:bar mass limit Trh}
\end{align}
where we have used~\cref{Eq:definition of reheating temperature}. {Additionally, $\upzeta$ is a dimensionless numerical factor, and we prefer to choose $\upzeta=10$ and $\upzeta=100$ in this work.} 

\begin{figure}[H]
    \centering
    \includegraphics[width=0.45\linewidth]{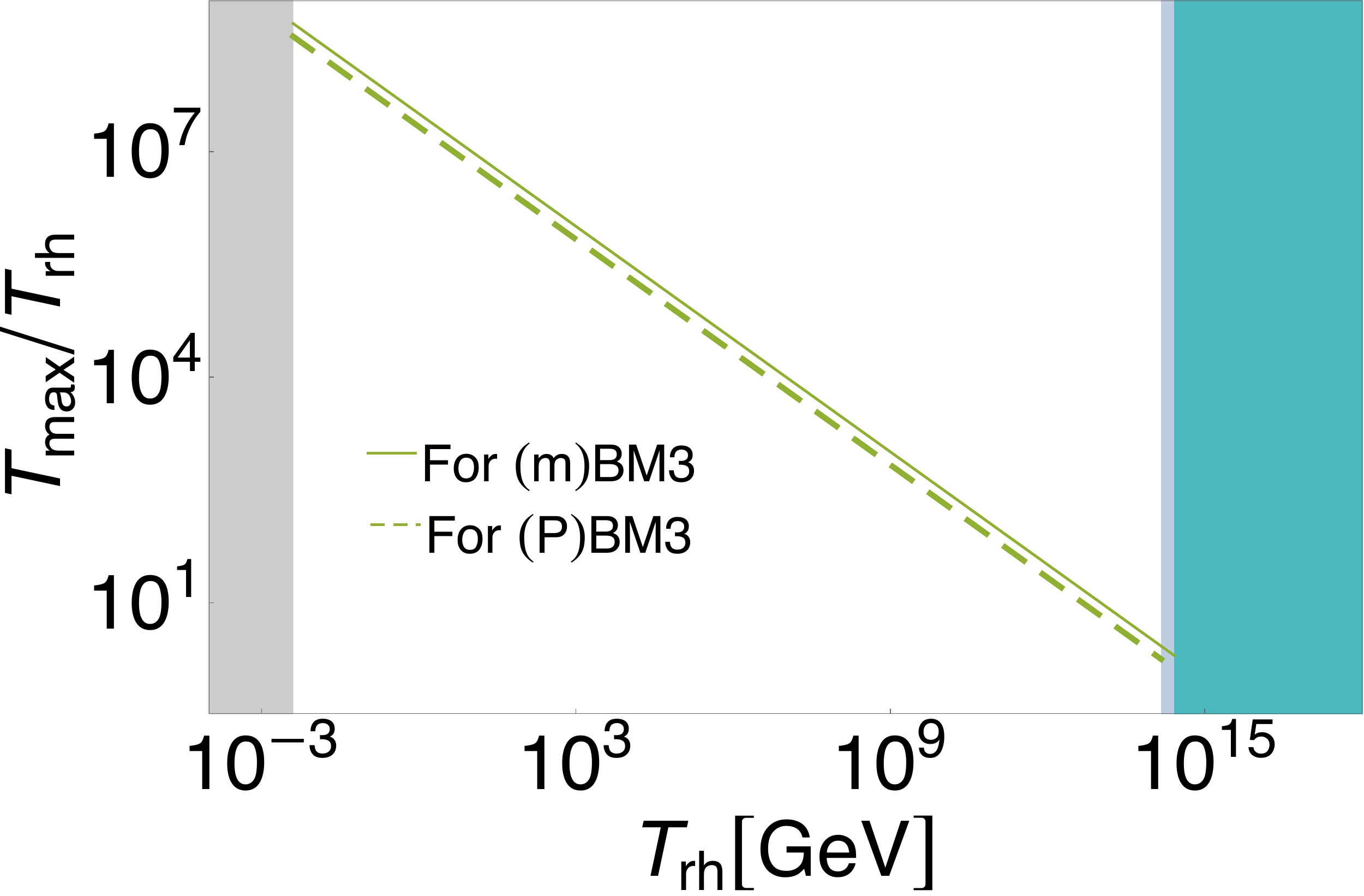}\;
\includegraphics[width=0.45\linewidth]{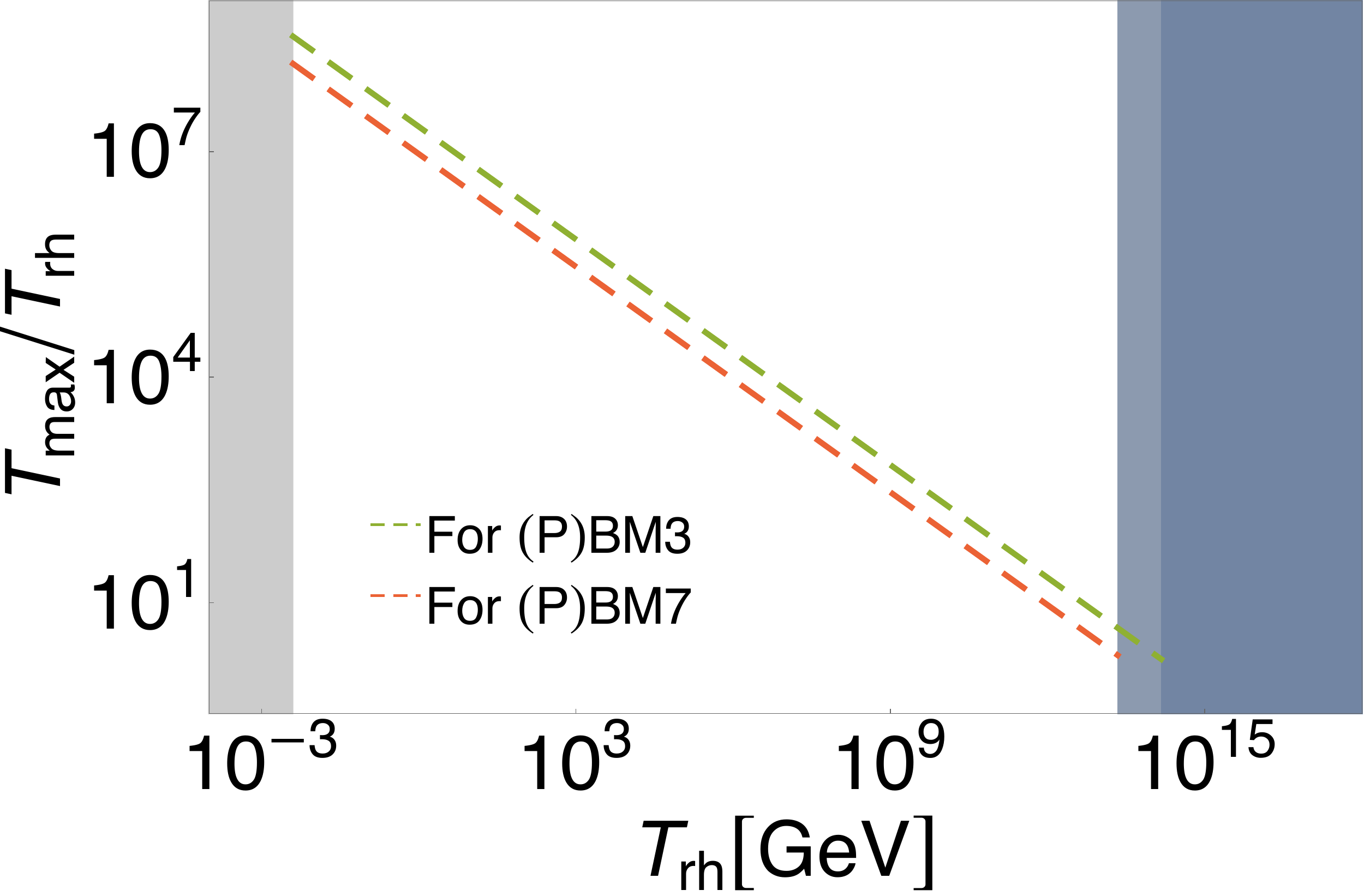}
    \caption{\it \raggedright
    Variation of $\Tmax/\Trh$ against $\Trh$. \textbf{Left panel:} shows $\Tmax/\Trh$ is less in Palatini formalism for same values of $\xi$ and $\ncmb$. \textbf{Right panel:} shows $\Tmax/\Trh$ decreases with increasing $\xi$ in Palatini formalism for same values of $\ncmb$. Solid line for metric formalism (actually for benchmark \mmoon), whereas dashed lines for Palatini formalism. The gray-colored stripe on the left denotes that $\Trh\nless 4\MeV$ are not allowed, and colored vertical stripes on the right 
    indicates those values of $\Trh$ are not allowed 
    from the stability analysis from~\cref{Table:Quartic_metric_stability,Table:Quartic_Palatini_stability}. 
    }
    \label{fig:Tmax-vs-Trh-plot}
\end{figure}

Now,~\cref{fig:Tmax-vs-Trh-plot} displays the variation of $\Tmax/\Trh$ against $\Trh$: left panel provides a comparative view of variation of $\Tmax/\Trh$ in metric (for benchmark value \mmoon) and Palatini formalisms (for benchmark value \Pmoon), for same values of $\upzeta=10$,  $\ncmb=60$ (and also for $\xi=10$). The gray-colored vertical stripe on the left represents the lower bound on $\Trh\gsim 4\times 10^{-3}\GeV$~\cite{Giudice:2000ex} (see also~\Ccite{Hasegawa:2019jsa,Kawasaki:2000en}). The colored vertical stripes on the right side represent those values of $\Trh$ are not allowed. The maximum allowed values of $\Trh$ correspond to the maximum allowed values of $\lO$ as mentioned in~\cref{Table:Quartic_metric_stability,Table:Quartic_Palatini_stability}
\footnote{These bounds correspond to ${\baremass}_{(m),\xi=10}$ and ${\baremass}_{{(P)},\xi=10}$ (for left panel) and ${\baremass}_{{(P)},\xi=10^4}$.}
. The bound for \mmoon~scenario is represented by the deep cyan stripe, whereas for~\Pmoon~scenario is shown by the deep blue (lighter tinted) region. From the left panel of~\cref{fig:Tmax-vs-Trh-plot} we see that $\Tmax/\Trh$ is greater in metric formalism (solid line) than in Palatini (dashed line) ($\Tmax/\Trh=5.0665\times 10^8$ for \mmoon, while $\Tmax/\Trh=3.3999\times 10^8$ for \Pmoon, at $\Trh=4\times 10^{-3}\GeV$) for the same values of $\xi$ and $\ncmb$. Right panel of~\cref{fig:Tmax-vs-Trh-plot} depicts $\Tmax/\Trh$ vs. $\Trh$ for two benchmark values in Palatini formalism - \Pmoon~and \Psun. It shows $\Tmax/\Trh$ decreases with higher values of $\xi$ ($\Tmax/\Trh=1.4736\times 10^8$ for \Psun, at $\Trh=4\times 10^{-3}\GeV$) which happens because potential becomes more flat for larger values of $\xi$. The same conclusion is also true in metric formalism, but the difference is not so much noticeable in comparison to Palatini formalism (e.g.  $\Tmax/\Trh=5.0665\times 10^8$ for \msun~at $\Trh=4\times 10^{-3}\GeV$). In the case of \Pmoon~scenario, the bound correspond to the maximum allowable value of $\Trh$ for  is denoted by the deep blue but lightly shaded vertical stripe, whereas, for \Psun~scenario, it is shown by the deep blue (darker shaded) region. To derive the maximum allowable value on $\Trh$, we have considered $\upzeta=10$ for both panels. For $\upzeta>10$, the maximum allowable value of $\Trh$ would increase further.


\subsection{Production of DM through decay}
\begin{figure}[H]
    \centering
    \includegraphics[width=0.45\linewidth]{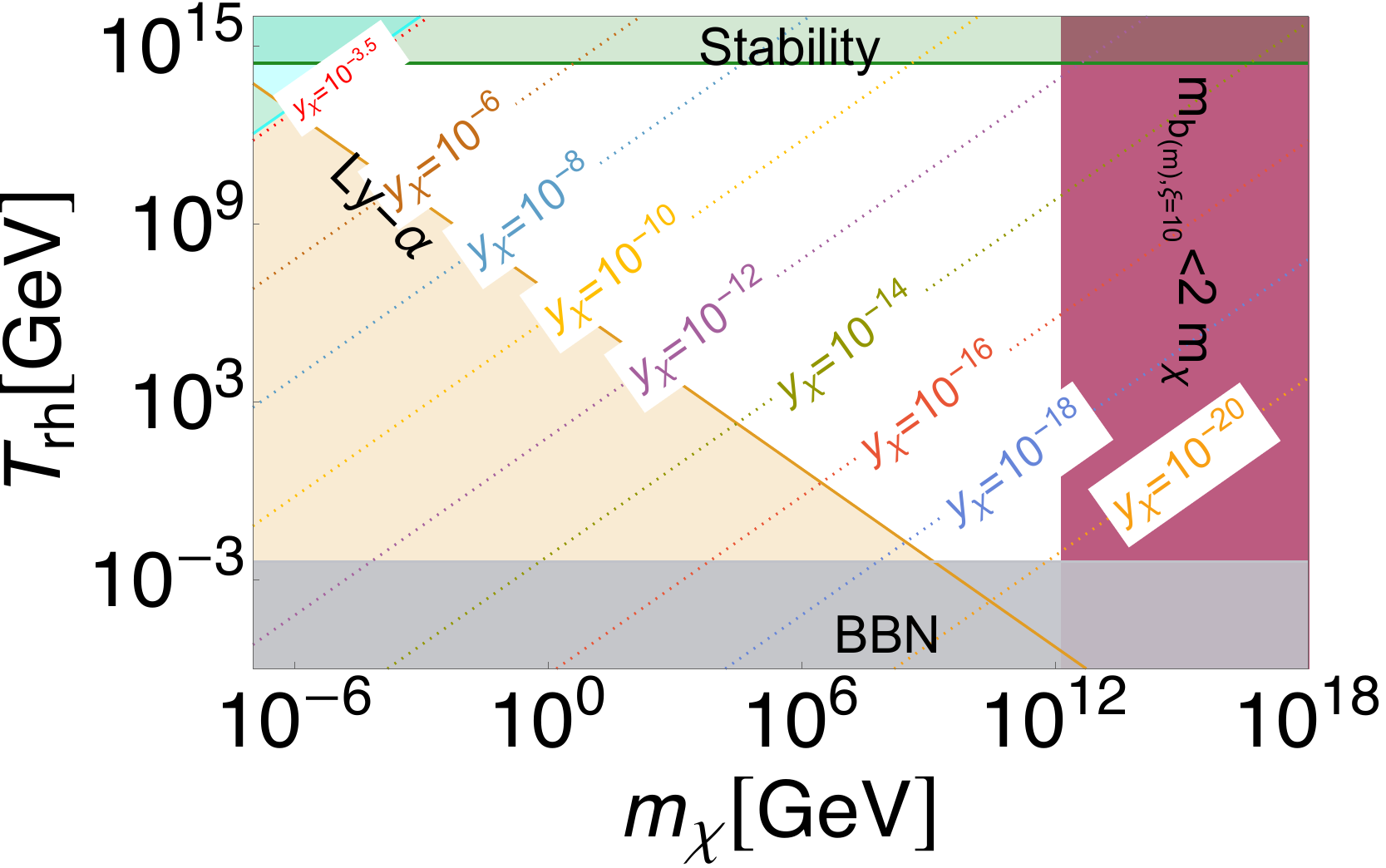}
    \includegraphics[width=0.45\linewidth]{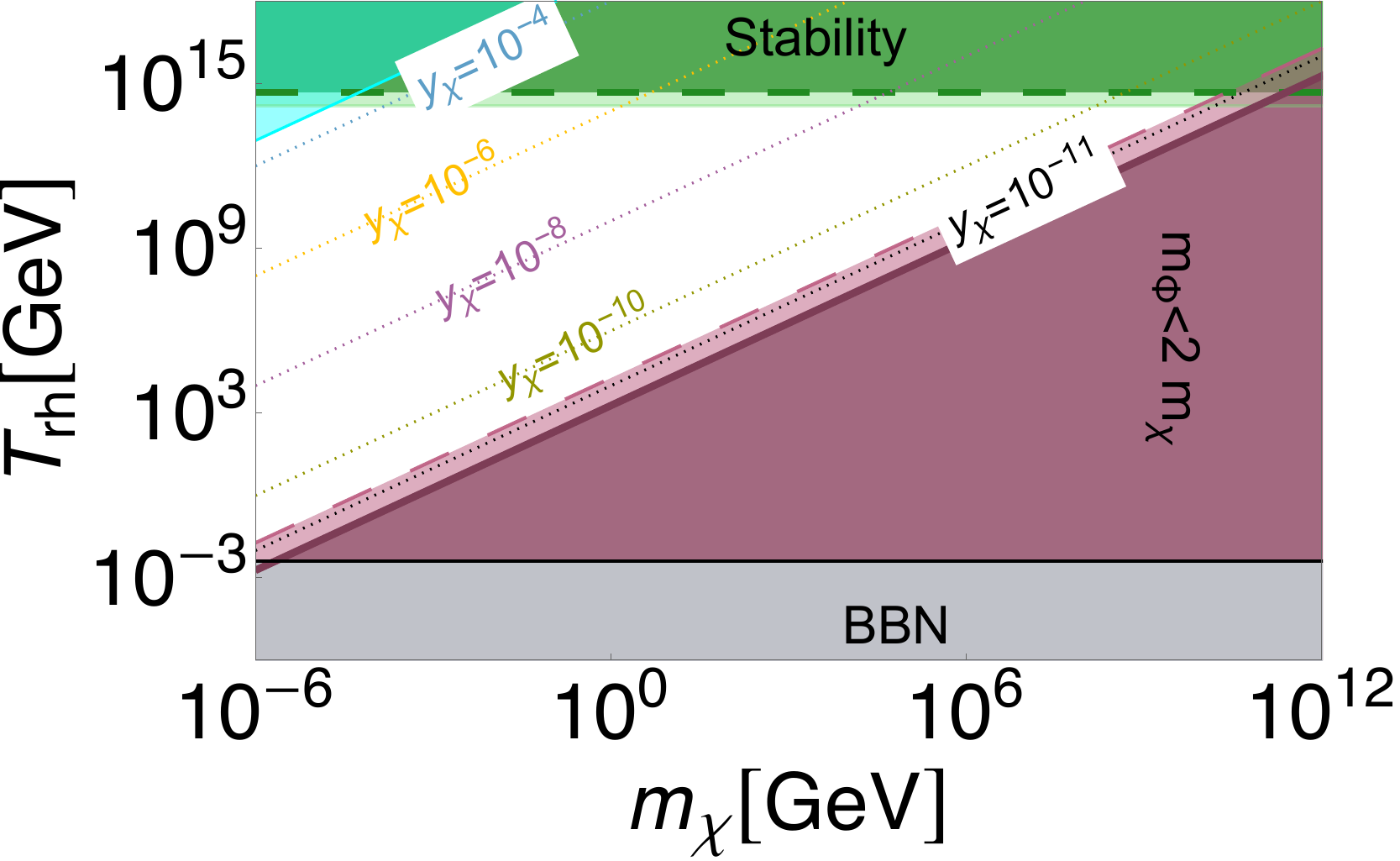}
    \includegraphics[width=0.45\linewidth]{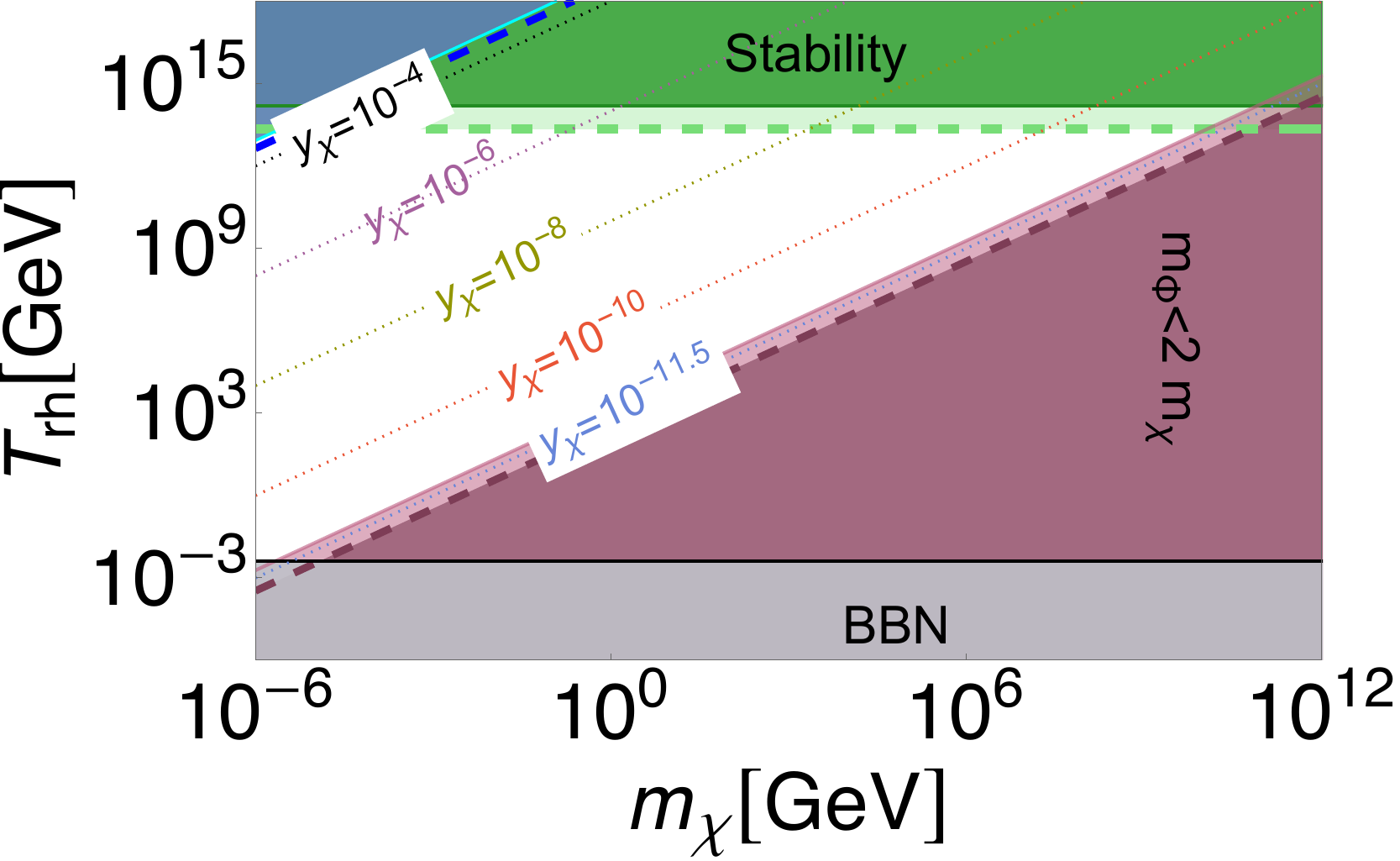}
    \caption{\it\raggedright The allowed region (unshaded area) on $(\Trh,\mc)$ plane and viable range for the Yukawa-like coupling $\yc$ to produce the entire \cdm~of the present universe: 
    {\bf top-left panel:} is for benchmark \mmoon, {\bf top-right panel:} is for benchmark \Pmoon~with bounds delimited by solid lines for $\upzeta=10$ and dashed lines for $\upzeta=100$, {\bf bottom panel:} is for benchmark \Pmoon~and \Psun~with bounds delimited by solid lines for $\xi=10$ and dashed lines for $\xi=10^4$, both for $\upzeta=10$. }
    \label{fig:Dm yield plot 2 metric+Palatini+Palatini}
\end{figure}

In this section, we investigate the case when the $\chi$-particles are produced from the decay of inflaton during the reheating era. We further assume that these $\chi$-particles, being both stable and non-thermal, act as viable \dm~candidates. Our aim is to explore the conditions under which these $\chi$-particles could potentially contribute to the entirety of the present-day \cdm~density in the universe. 
If $n_\chi$ and $N_\chi$ denotes the number density and comoving number density of \dm~particles, then
the evolution equation of comoving number density of $\chi$ is
\begin{align}\label{eq:Comoving dm number density}
    \frac{\dd N_\chi}{\dd t} = \cs^3 \, \gamma\,,
\end{align}
where $\gamma$ is the rate of production of $n_\chi$ and for $(\phi\to \cc)$ 
\begin{align}\label{eq:gamma defined}
    \gamma= 2 \, \Br \, \Gamma_\phi \,\frac{\rho_\phi}{m_\phi}\,.
\end{align}
Here, '$\Br$' denotes branching fraction for the production of $\chi$ from the decay process, and using~\cref{Eq:decay-width-of-inflaton,eq:decay width}, it can be expressed as~\cite{Ghoshal:2023jhh,Ghoshal:2023jvf} 
\begin{align}\label{eq:Br defined}
    \Br=\frac{\Gamma_{\phi\to \cc}}{\Gamma_\phi}\approx m_\phi^2 \lt( \frac{\yc}{\lO}\rt)^2\,.
\end{align}
Following~\cref{appendix:Boltzmann},
for $\Tmax>T>\Trh$, $\hubble$ and $\r_\phi$  can be defined as~\cite{Bernal:2021qrl}
\eq{
\label{Eq:Hubble parameter during reheating+Eq:rho_phi}
{\cal H} \simeq \frac{\pi}{3} \sqrt{\frac{\gs}{10}} \frac{T^4}{M_P\, \Trh^2}\,,
\qquad 
&\r_{\phi} \simeq \frac{ \pi^2  \gs }{30 } \frac{T^8}{ \Trh^4} \,.
}
%
Using~\cref{eq:gamma defined,eq:Br defined,Eq:Hubble parameter during reheating+Eq:rho_phi,eq:Last pole} in~\cref{eq:Comoving dm number density}, we can obtain \dm~yield, $\Yc$ which is defined as the ratio of number density of $\chi$ to entropy density, as 
\begin{empheq}[
  left=
    {
 \Yc\simeq 
} 
    \empheqlbrace
]{align}\label{eq:Yield of DM-metric}
   & \frac{3}{\pi} \frac{\gs}{\gss} \sqrt{\frac{10}{\gs}} \frac{\mpl \, \Gamma_\phi}{m_{\phi}\, \Trh} \text{Br} \, \quad \text{(in metric formalism)}\,,\\%
 %
 %
 \label{eq:Yield of DM-Palatini}
 &\frac{9}{2\pi} \frac{\gs}{\gss} \sqrt{\frac{10}{\gs}} \frac{\mpl \, \Gamma_\phi}{m_{\phi}\, \Trh} \text{Br}\, \quad \text{(in Palatini formalism)}\,.
\end{empheq}
Present day \cdm~yield is given by
\eq{\label{Eq:present day CDM yield}
 Y_{{\rm CDM},0}  
 =\frac{4.3. \times 10^{-10}}{\mc/\GeV} \,.
}
The condition $\Yc\sim  Y_{{\rm CDM},0}$  leads to
\begin{empheq}[
  left=
    {
 \Trh\seq
} 
    \empheqlbrace
]{align}
& 6.4907 \times 10^{25}\, \yc^2 \, \mc \,\quad \text{(in metric formalism)} \,, \label{eq:DM yield line m}\\%
& 9.7360 \times 10^{25}\, \yc^2 \, \mc \,\quad \text{(in Palatini formalism)} \,. \label{eq:DM yield line P}
\end{empheq}
If \dm~particles are produced from the decay of inflaton during reheating era in our considered inflationary scenario, and 
account for $100\%$ of the total CDM density of the present universe, then it must satisfy~\cref{eq:DM yield line m,eq:DM yield line P}. These equations are represented as dotted lines for various fixed values of $\yc$ on $(\Trh, \mc)$ plane, as shown in~\cref{fig:Dm yield plot 2 metric+Palatini+Palatini}. Top-left panel of this figure for metric formalism (for \mmoon~with $\upzeta=10$), top-right panel for Palatini formalism (for \Pmoon~with $\upzeta=10$ and $\upzeta=100$), and bottom panel for Palatini formalism but for \Pmoon~and \Psun~with $\upzeta=10$.  The bounds on $(\Trh,\mc)$ plane are mentioned below. 

\begin{itemize}
    \item {\bf The maximum permissible value $\Trh$:} This bound corresponds to the maximum permissible value of $\Trh$ which in turn correspond to the maximum allowed value of $\lO=\lambda_{12,{\rm max}}$ from~\cref{Table:Quartic_metric_stability,Table:Quartic_Palatini_stability}. This bound can be obtained from~\cref{Eq:definition of reheating temperature,eq:baremass-metric,eq:bar mass limit Trh}, and it is as follows:
\begin{empheq}[
  left=
    {
\Trh\Big|_{\rm max} \simeq
} 
    \empheqlbrace
]{align}
&\sqrt{\frac{2}{\pi}} \left(\frac{10}{\gs}\right)^{1/4} \sqrt{\mpl} \lt(\frac{\lambda_{12,{\rm max}}^2}{8\pi\, {\baremass}_{(m),\xi}} \rt)^{1/2} \, \quad &\text{(for metric)}\,,\\
&\sqrt{\frac{\upzeta}{2}}  
         \lt(\frac{1}{3 \, \pi^{10}\,\Lambda} \rt)^{1/12}
         \left(\frac{10}{\gs}\right)^{1/4} {\mpl}^{1/3} {\lambda_{12,{\rm max}}}^{2/3}     \, \quad &\text{(for Palatini)}\,.
\end{empheq}
This is displayed as horizontal forest-green or pastel-green shaded horizontal stripe placed at the upper side of the plots.
\item {\bf The maximum value of $\yc$:} This bound corresponds to the maximum permissible value of $\yc={\yc}_{,{\rm max}}$ obtained from~\cref{Table:Quartic_Palatini_stability,Table:Quartic_metric_stability}. This bound varies in metric and Palatini formalism as outlined below (from~\cref{eq:DM yield line m,eq:DM yield line P}): 
\begin{empheq}[
  left=
    {
 \Trh\seq
} 
    \empheqlbrace
]{align}
& 6.49 \times 10^{25}\, {\yc}_{\rm max}^2 \, \mc \,\quad \text{(in metric formalism)}  \,,\\%
& 9.74 \times 10^{25}\, {\yc}_{\rm max}^2 \, \mc \,\quad \text{(in Palatini formalism)}  \,.
\end{empheq}
This is displayed as cyan colored (or deep blue colored on the plot at the bottom plane) wedge-shaped region at the left-corner of the plots. 
\item {\bf The maximum possible value of $\mc$ in decay process:} which is $\mc\lsim {\baremass}_{(m),\xi}/2$ (in metric formalism), and $\mc\lsim {\baremass}_{(P),\xi}/2$ (in Palatini formalism). In metric formalism, the region for $\mc> {\baremass}_{(m),\xi}/2$  is indicated by a vertical copper-rose stripe on the right-side of the plot. 
However, in the Palatini case, due to the dependence of ${\baremass}_{(P),\xi}$ on $\Trh$, the representation for the region $\mc> {\baremass}_{(P),\xi}/2$ changes to a wedge-shaped copper-rose region instead of a vertical stripe. 

\item {\bf Lyman-$\alpha$ bound:} This constraint guarantees that $\chi$-particles do not contribute to warm dark matter, but rather to \cdm. Since the $\chi$-particles are feebly interacting, their velocity can only decrease due to redshift.
Based on this the lower bound on $\mc$, as derived in~\Ccite{Bernal:2021qrl} (see also discussion in~\Ccite{Ghoshal:2022jeo}), is 
\begin{align}\label{eq:Lyman-alpha}
        {\mc} \gsim 2 \times 10^{-6}\frac{m_\phi}{\Trh}\,\GeV\,.
    \end{align}
In top-left panel (for the case of metric formalism) it is denoted by the peach-colored wedge-shaped region at the left bottom of the plot. 
%
However, in Palatini formalism, this bound can be obtained from~\cref{eq:bar mass limit Trh} as
    \eq{
\mc\gsim& \frac{2 \times 10^{-6}}{\upzeta} \,  \lt(\frac{3 \pi^2}{40 }\rt)^{1/4}  \, \qty(\gs \, \Lambda)^{1/4} \,\GeV\,\\
\gsim&
\begin{cases}
    2.1596\times 10^{-9} \GeV\, \quad &(\text{for (P)BM3 with} \, \upzeta=10, \Lambda=  1.7206\times 10^{-10}) \,,\\
    2.1596\times 10^{-10} \GeV\, \quad &(\text{for (P)BM3 with} \, \upzeta=100, \Lambda=  1.7206\times 10^{-10}) \,,\\
    1.2154\times 10^{-8} \GeV\, \quad &(\text{for (P)BM7 with} \, \upzeta=10, \Lambda=  1.7261\times 10^{-7})\,.
\end{cases}
}
{Since this bound is very small for Palatini case,  it is omitted from the top-right and bottom panels of~\cref{fig:Dm yield plot 2 metric+Palatini+Palatini}. Actually, the lower bound on warm dark matter particle mass is set at $\gsim 3.5 \unit{keV}$~\cite{Bernal:2021qrl}, rising to $\gsim 5.3 \unit{keV}$~\cite{Viel:2013fqw,Palanque-Delabrouille:2019iyz} at $95\%$ CL for thermal warm dark matter. Uncertainties in the thermal history of the universe may reduce this to $1.9 \unit{keV}$~\cite{Garzilli:2019qki}. For other Feebly Interacting Massive Particles are bounded between $4$ and $16 \unit{keV}$~\cite{Bae:2017dpt,Murgia:2017lwo,Heeck:2017xbu,Boulebnane:2017fxw,Baldes:2020nuv,Ballesteros:2020adh,DEramo:2020gpr,Decant:2021mhj}.
}

\item {\bf BBN :} $ T_{rh}\gsim 4 \MeV$. This is presented as a horizontal slate-gray stripe at the bottom of the plots. 
 
\end{itemize}
The unshaded region in the plot at top-left panel of~\cref{fig:Dm yield plot 2 metric+Palatini+Palatini} represents the permissible parameter space on $(\Trh,\mc)$ plane for \mmoon~with $\upzeta=10$. Since, there is no significant difference in  ${\baremass}_{(m),\xi=10}$ (for \mmoon) and in ${\baremass}_{(m),\xi=10^4}$ (for \msun), only the allowed region for \mmoon~is shown here (${\baremass}_{(m),\xi=10}=1.1307\times 10^{-6}\mpl$, ${\baremass}_{(m),\xi=10^4}=1.1212\times 10^{-6}\mpl$ for $\upzeta=10$). 
Similarly, the bottom panel of~\cref{fig:Dm yield plot 2 metric+Palatini+Palatini} presents the permissible parameter region on $(\Trh,\mc)$ plane for \Pmoon~and \Psun~with $\upzeta=10$: Continuous lines demarcate bounds corresponding to \Pmoon, while dashed lines demarcate bounds corresponding to \Psun. 
Analysis of this figure leads to the conclusion that $\chi$ particles, produced from the decay of inflaton for the case of benchmark \mmoon, can explain the total CDM density, provided that $\yc$ lies within the range $10^{-3.5}\gsim \yc\gsim 10^{-20}$, and $\mc$ fall within $7.1008\times 10^{-7}\lsim \mc /\GeV\lsim {\baremass}_{(m),\xi=10}$). Conversely, in Palatini formalism, the upper limit varies with $\Trh$. Additionally, lower limit of $\yc$ is $\sim 10^{-11}$, which is higher in comparison to metric case. With higher value of $\upzeta$,  the allowable range of $\yc$ narrows (e.g. see dotted line correspond to $\yc =10^{-11}$), and  broadens with larger values of $\xi$.

The benchmarks we have chosen in metric formalism for $\ncmb\sim 60$, mentioned in~\cref{table:quartic-Benchmark-metric}, can be validated in the future by the forthcoming \cmb~observations.
These future observations, therefore, can help in verifying whether the range of $10^{-11}\gsim \yc \gsim 10^{-20}$ is valid.  Furthermore, the forthcoming \cmbsfour~observation can validate, as we have discussed earlier, the benchmarks we have chosen in the Palatini formalism. As an illustrative example, \cmbsfour~can affirm the validity of benchmark (P)BM3, thereby aiding in determining that a range around $\yc\sim 10^{-11.5}$ is valid for a fixed value of $\upzeta$, let's say $\upzeta=10$. In summary, these forthcoming CMB observations should be able to provide the validity  of the benchmarks we have chosen, both in metric and Palatini formalisms, along with the valid range of values of $\yc$.




\section{non-thermal leptogenesis}
\label{sec:lepto}
%


Here, we consider the production of non-thermal right-handed heavy sterile (i.e. fermionic fields which are gauge singlet under \sm~gauge groups) neutrinos, $N_{j}$ during the reheating era from the decay of the inflaton. Then the Lagrangian density is given by~\cite{Fong:2012buy,Rubakov:2017xzr,DiBari:2012fz,Buchmuller:2005eh,Davidson:2008bu,Trodden:2004mj} 
\begin{align}
    {\cal L}_{\rm lepto}=&{i}\sum_{j=1,2,3}\bar{N}_{j}\gamma_\mu\partial^\mu {N}_{j} - \frac{\MN_j}{2} \bar{N}^c_{j} N_{j} 
    -  \sum_{a,b,j=1,2,3}\lt(\epsilon_{a b} \, \Yu_{d j}\,
 \bar{N}_{j}\, {L}^a_{d} \, H^b    \,+\, \text{h.c.}\rt) +  {\cal L}_{\phi\to N\bar{N}}+  {\cal L}_{\phi\to \HH}\,,\label{eq:Lepto-lagrangian}
\end{align}
where 
$\gamma_{\mu}$ are the gamma matrices, $\bar{N}_{j}=N_{j}^\dagger \gamma^0$, with superscript $c$ indicating charge conjugation,
$\epsilon_{ab}=i\sigma_2$ is the Levi-Civita symbol in 2-D and $\sigma_2$ is the Pauli matrix, Latin index $d$ corresponds to three generations-$e,\mu,$ and $\tau$. In~\cref{eq:Lepto-lagrangian}, $\MN_j$s represents the  Majorana masses of the sterile neutrinos, forming a real diagonal mass matrix in the chosen basis,
whereas $\Yu_{dj}$ is a complex number denoting the Yukawa coupling matrix. 
The presence of a non-zero value for $\MN_j$ leads to the violation of lepton number, fulfilling Sakharov's first condition, while the complex $\Yu_{dj}$ is essential for CP violation, satisfying Sakharov's second condition to generate baryon asymmetry~\cite{Fong:2012buy}.
Additionally, 
$L_d$ is the left leptonic doublet, defined as $L_d\equiv (\nu_{d}, d_l^-)$, where $\nu_{d}$ are the light left-handed neutrinos in \sm~\cite{Fong:2012buy}. 
The ‘h.c.’ in~\cref{eq:Lepto-lagrangian} denotes hermitian conjugated of interaction terms $\sim   \Yu^*_{d j} \, \tilde{H}^b \,\bar{L}^a_{d}\,  {N}^c_{j}$, where $\tilde{H}_b=\epsilon_{b a}H^{\dagger a}$. ${\cal L}_{\phi\to N_j\bar{N}_j}=-y_{N_j}\,  \phi \, \bar{N}_j N_j$, and ${\cal L}_{\phi\to \HH}$ in~\cref{eq:Lepto-lagrangian}  indicates that \sm~Higgs particles are also produced along with $N_j$ from the inflaton decay during the reheating era, and it is defined as
  \begin{align}
    {\cal L}_{\phi\to \HH}=  - \lO \, \phi \, \HH 
- \lT \, \phi^2 \, \HH 
+ \text{h.c.}\,.
  \end{align}
Similar to~\cref{Eq:decay-width-of-inflaton},  in this case, we also assume that the dominant decay channel is to \sm~Higgs and as a result, $\Trh$ is still determined by~\cref{Eq:definition of reheating temperature}. Actually, $\chi$ is very similar to $N_j$, as they both share the feature that when the \bsm~particle $\chi$ is stable, it contributes to \cdm~energy density of the present universe. However, when $\chi$ is not stable and decays to \sm~leptons, it plays a role in producing the observable baryon asymmetry in the universe. 
However, left- and right-handed vector-like fermions transform identically~\cite{delAguila:1989rq}. Therefore, decay width for $\phi\to N_j\bar{N}_j$ is 
\eq{\label{eq:Gamma:phi-to-NN}
 \Gamma_{\phi \to N_j N_j}  \simeq \frac{y_{N_j}^2\, m_{\phi}}{4\pi}\,,
}
where $N_j$  is the sterile neutrino particle of $j$-th generation.

Due to non-zero value of $M_j$, $N_j$ can decay via two decay channels: $N_j \rightarrow L_d +H$, and $ N_j \rightarrow \bar{L}_d + \tilde{H}$~\cite{Co:2022bgh,Fong:2012buy,Asaka:1999yd}. 
If $\Gamma_j(N_j \rightarrow L_d +H)$ and $\Gamma_j(N_j \rightarrow \bar{L}_d + \tilde{H})$ are respective decay width of those decay channels, then Sakharov's third condition of thermal inequilibrium is satisfied if $\Gamma_j(N_j \rightarrow L_d +H) + \Gamma_j(N_j \rightarrow \bar{L}_d + \tilde{H}) \gg \hubble$, and from~\cref{sec:reheating} we see that it happens when $T$ just reach and drops below $\Trh$. In this work, we assume $N_1$ is the lightest among the three sterile neutrinos. Hereafter, we focus on the asymmetry generated by the decay of $N_1$, as asymmetry produced in the decays of heavier particles gets washed out by the asymmetry generated by the lightest ones~\cite{Rubakov:2017xzr}. Considering both tree-level and one-loop diagrams, the lepton asymmetry generated by the decay process of $N_1$ is expressed as~\cite{Asaka:1999yd,Fukuyama:2005us,Hamaguchi:2002vc}
\begin{align}\label{eq:epsilon-lepto-def}
    \epsilon^{\lep}_1= \frac{\Gamma_1(N_1 \rightarrow L_d +H) - \Gamma_1(N_1 \rightarrow \bar{L}_d + \tilde{H})}%
    {\Gamma_1(N_1 \rightarrow L_d +H) + \Gamma_1(N_1 \rightarrow \bar{L}_d + \tilde{H})}\,.
\end{align}
By combining contributions from one-loop vertex and self-energy corrections,~\cref{eq:epsilon-lepto-def} leads to~\cite{SravanKumar:2018tgk,Fukuyama:2005us,Asaka:1999yd}
\begin{equation}
\epsilon^{\lep}_{1}=-\frac{1}{8\pi}\frac{1}{\left(\Yu \, \Yu^{\dagger}\right)_{11}}\sum_{j=2,3}\text{Im}\left[\left\{ \left(\Yu \, \Yu^{\dagger}\right)_{1j}\right\} ^{2}\right]\left[f\left(\frac{{\MN_j}^{2}}{{\MN_1}^{2}}\right)+ 2g\left(\frac{{\MN_j}^{2}}{{\MN_1}^{2}}\right)\right]\,.\label{epsilon_i-exp}
\end{equation}
The functions $f(x)$ and $g(x)$, where $x={{\MN_j}^{2}}/{{\MN_1}^{2}} $, accounts the contributions from vertex and self-energy corrections/wave function corrections~\cite{Hamaguchi:2002vc,Fukuyama:2005us}, and they are given by~\cite{SravanKumar:2018tgk,Hamaguchi:2002vc}
\begin{align}  f\left(x\right)=\sqrt{x}\left[-1+\left(x+1\right)\ln\left(1+\frac{1}{x}\right)\right]\,,\quad g\left(x\right)=\frac{\sqrt{x}}{2(x-1)}\,.
\end{align}
Under the assumption of mass hierarchy $\MN_1\ll \MN_{2}, \MN_3$ i.e. $x\gg 1$~\cite{Hamaguchi:2002vc}, $f(x)\sim \frac{1}{\sqrt{2}\, x}$ and $g(x)\sim \frac{1}{2\,\sqrt{ \, x}}$~\cite{Fukuyama:2005us}. Then, we obtain from~\cref{epsilon_i-exp}
\begin{align}\label{eq:epsilon-step-1}
    \epsilon^{\lep}_1=-\frac{3}{16 \pi}\, \frac{1}{\left(\Yu \, \Yu^{\dagger}\right)_{11}}\sum_{\mathfrak{i}=2,3}\text{Im}\left[\left\{ \left(\Yu \, \Yu^{\dagger}\right)_{1\mathfrak{i}}\right\} ^{2}\right] \, \frac{\MN_1}{\MN_j}\,.
\end{align}
Using $\Im{\lt(\qty(\Yu\Yu^\dagger)_{11}\rt)^2}=0$ and properties of inverse of a diagonal matrix, we can rewrite~\cref{eq:epsilon-step-1} as~\cite{Hamaguchi:2002vc}
\begin{align}
   \epsilon^{\lep}_1= -\frac{3}{16 \pi}\, \frac{\MN_1}{\left(\Yu \, \Yu^{\dagger}\right)_{11}}%
   \Im{\lt(\Yu \Yu^\dagger \frac{1}{\mathbb{M}}\Yu^* \Yu^T\rt)_{11}}\,,
\end{align}
where $\mathbb{M}$ is the diagonal Majorana mass matrix. To relate lepton asymmetry with the mass of $j-$th generation \sm~light neutrino, $m_{\nu_j}$, we use Type-I see-saw mass of the neutrino $\mathbb{m}_\nu = - \Yu^T \, \mathbb{M}^{-1}\, \Yu \expval{H}$, where $\mathbb{m}_\nu$ is the mass matrix 
of \sm~neutrinos and  $\expval{H}=246/\sqrt{2}\GeV$~\cite{Fukuyama:2005us,Hamaguchi:2002vc}. 
$\mathbb{m}_\nu$ is 
diagonalizable matrix, i.e. it can be diagonalized by unitary transformation as~\cite{Hamaguchi:2002vc}
\eq{
\lt( \mathbb{m}_\nu \rt)_{\alpha \beta} = \sum_{\mathfrak{i} } U_{\alpha \mathfrak{i}} U_{\beta \mathfrak{i} } \lt( \widehat{\mathbb{m}_\nu} \rt)_{\mathfrak{i}}\,.
}
Here, $\widehat{\mathbb{m}_\nu}$ represents the diagonalized mass matrix $\widehat{\mathbb{m}_\nu} = \text{diag}\lt({m_{\nu}}_1, {m_{\nu}}_2, {m_{\nu}}_3 \rt)$ with ${m_{\nu}}_1< {m_{\nu}}_2< {m_{\nu}}_3$. Similarly, we can define $\widehat{\Yu}$ as
\eq{
\widehat{\Yu}_{\mathfrak{i}j} = \sum_\alpha  \widehat{\Yu}_{\mathfrak{i}\alpha} U^*_{\alpha j}\,.
}
Consequently, we obtain~\cite{Asaka:1999yd}
\begin{align}
    \epsilon^{\lep}_1\seq \frac{3}{16 \pi}\, \frac{\MN_1}{\expval{H}^2} \,  m_{\nu_3} \, \delta_{eff}\,,
\end{align}
where $\delta_{eff}$ is the measure of effective CP violating phase, which is defined as~\cite{Hamaguchi:2002vc}
\eq{
\delta_{eff}= \frac{\Im \lt[ \lt(\widehat{\Yu}_{13}\rt)^2 + \frac{{m_{\nu}}_2}{{m_{\nu}}_3}\lt(\widehat{\Yu}_{12}\rt)^2 +\frac{{m_{\nu}}_1}{{m_{\nu}}_3}\lt(\widehat{\Yu}_{11}\rt)^2 \rt]}{ \lt|\widehat{\Yu}_{13}\rt|^2 + \lt|\widehat{\Yu}_{12}\rt|^2+ \lt|\widehat{\Yu}_{11}\rt|^2}\,,
}
and $0\leq \delta_{eff}\leq 1$~\cite{Co:2022bgh}. If we assume that produced $N_1$ decays immediately after production from inflaton decay, then produced Lepton asymmetry is expressed in terms of lepton-to-entropy ratio as~\cite{Hamaguchi:2002vc}
\begin{align}\label{eq:Lepton asymmetry}
    Y_L\equiv \frac{n_L}{s}= \epsilon^{\lep}_1 \frac{n_{N_1}}{s}\,,
\end{align}
where $n_L$ is the number density of excess Leptons produced due to decay of $N_1$ neutrino and $n_{N_1}$ is the number density of $N_1$ neutrino produced from the decay of inflaton. 
The value of ${n_{N_1}}/{s}$ can be obtained by utilizing~\cref{eq:Yield of DM-metric,eq:Yield of DM-Palatini,eq:Gamma:phi-to-NN} as
\begin{empheq}[
  left=
    {
 \frac{n_{N_1}}{s}\seq
} 
    \empheqlbrace
]{align}
& 2.3258 \times 10^{-2}\, \frac{\mpl}{\Trh}\,{y_{N_1}}^2 \,\quad &\text{(in metric formalism)}\,,  \\%
& 3.4887 \times 10^{-2}\,\frac{\mpl}{\Trh}\, {y_{N_1}}^2 \, \quad &\text{(in Palatini formalism)}  \,.
\end{empheq}

As a result of electroweak sphaleron processes, the lepton asymmetry undergoes conversion into the baryon asymmetry. Hence, the baryon asymmetry is proportional to the lepton asymmetry, leading to~\cite{SravanKumar:2018tgk,Co:2022bgh}
\begin{align}
    Y_B=\frac{28}{79} Y_L\,.
\end{align}



\begin{figure}[H]
    \centering
    \includegraphics[width=0.5\linewidth]{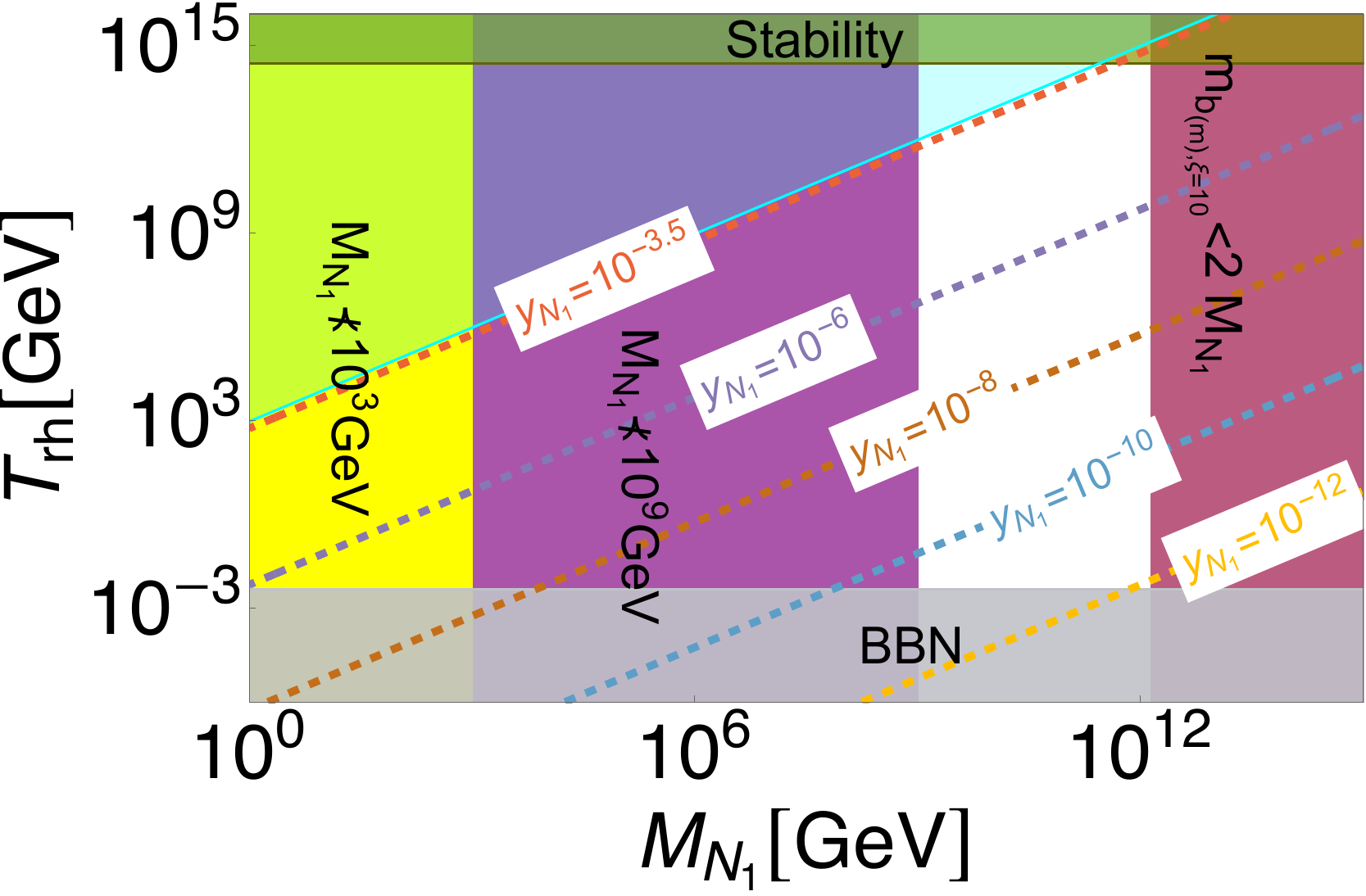}
    \caption{\it \raggedright
    The allowed region (unshaded area) on $(\Trh,M_{N_1})$ plane and viable range for the Yukawa-like coupling $y_{N_1}$ to produce the baryon asymmetry of the present universe for \nmpq~in metric formalism for the benchmark \mmoon. 
    }
    \label{fig:metric-lepto}
\end{figure}


 The value of $Y_B$ from \Planck2018 data can be obtained as~\cite{Rubakov:2017xzr,Co:2022bgh}
\begin{align}\label{eq:YB bound}
    Y_B\equiv \frac{n_B- n_{\bar{B}}}{s}= \frac{n_\gamma}{s}\eta_B  \seq 8.7\times 10^{-11}\,,
\end{align}
where $n_B- n_{\bar{B}}$, $n_\gamma$, and $\eta_B$ are the difference in the number density of baryons and anti-baryons, the number density of \cmb~photons, and baryon-to-photon ratio, respectively. In~\cref{eq:YB bound}, we use the values of present day number density of photons $n_\gamma\sim 410.73 \, \unit{cm}^{-3}$, and present day value of entropy density $s=2891.2 \, \unit{cm}^{-3}$ (as we are using $k_B=1$ unit) obtained from \Planck2018 data.
Utilizing~\cref{eq:Lepton asymmetry,eq:YB bound}, we get for both metric and Palatini formalisms as 
\eq{\label{eq:lepto-condition}
\Trh\seq
& 5.6073 \times 10^{9}\, y_{N_1}^2 \, \MN_1  
\,.
%
}
Here, we consider $\delta_{eff}=0.5$, and $m_{\nu_3} \approx 0.05\times 10^{-9}\GeV$. 
Thus, if the sterile neutrino is produced from the decay of inflaton during the reheating era, and becomes accountable for the generation of baryon asymmetry of the present universe, then it must satisfy~\cref{eq:lepto-condition}. 
These are presented as dotted lines for different values of $y_{N_1}$ in~\cref{fig:metric-lepto}. 
In this plot, the unshaded region represents the permissible parameter space on $(\Trh, M_{N_1})$ plane for benchmark \mmoon. Constraints on this plane include: 
a forest-green horizontal stripe situated at the top of the plot, representing the maximum permissible $\Trh$ value; a cyan-colored wedge-shaped region in the upper-left corner. Both of these  bounds are derived from~\cref{Table:Quartic_metric_stability}. Additionally, a copper-rose vertical stripe on the right side reflects the condition $M_{N_1}$can not be $> {\baremass}_{(m),\xi}/2$, while a slate-gray horizontal stripe at the bottom rules out the possibility $\Trh< 4 \MeV$. These constraints have been discussed in the context of~\cref{fig:Dm yield plot 2 metric+Palatini+Palatini}. Furthermore, purple-colored region excludes $M_{N_1}<10^9\GeV$~\cite{Davidson:2002qv}, and a yellow-colored region similarly eliminates $M_{N_1}<10^3\GeV$. 
Hence, for $10^{-3.5}\gsim y_{N_1}\gsim 10^{-12}$, the sterile neutrino produced from the decay of inflaton within the metric formalism of our considered inflationary scenario can effectively account for the entire baryon asymmetry in the present universe.

\begin{figure}
    \centering
    \includegraphics[width=0.45\linewidth]{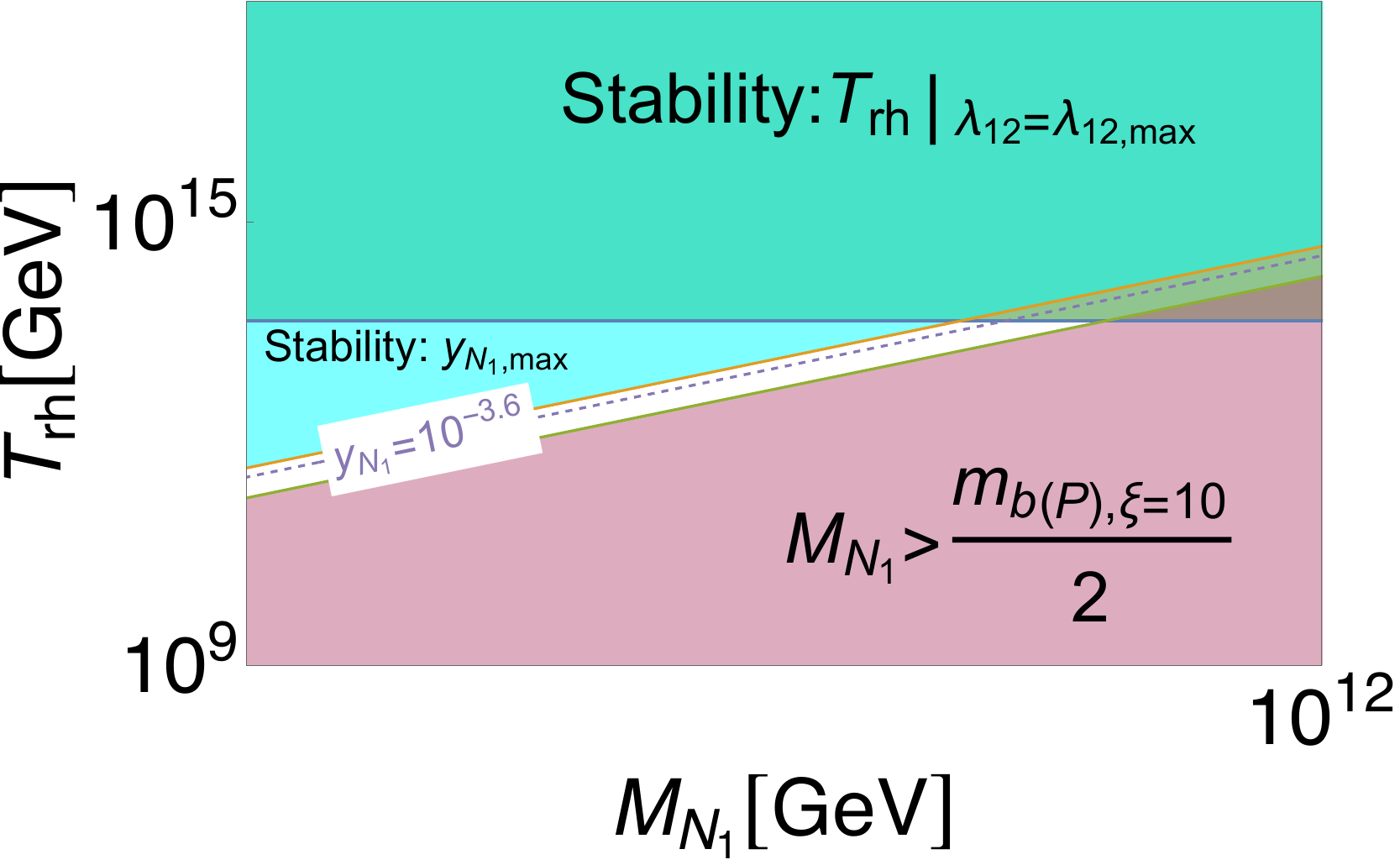}\;
    \includegraphics[width=0.45\linewidth]{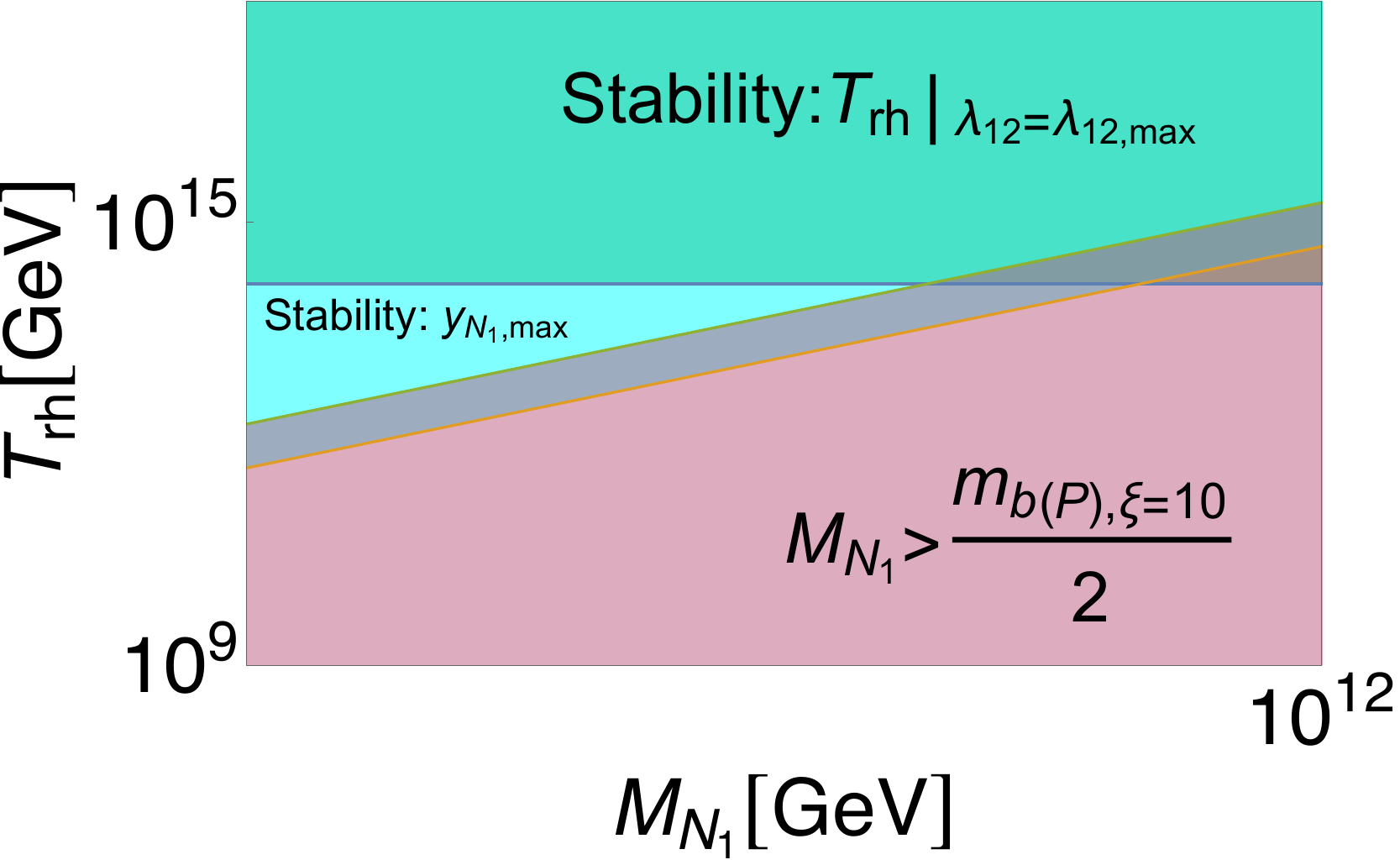}\;
    \includegraphics[width=0.45\linewidth]{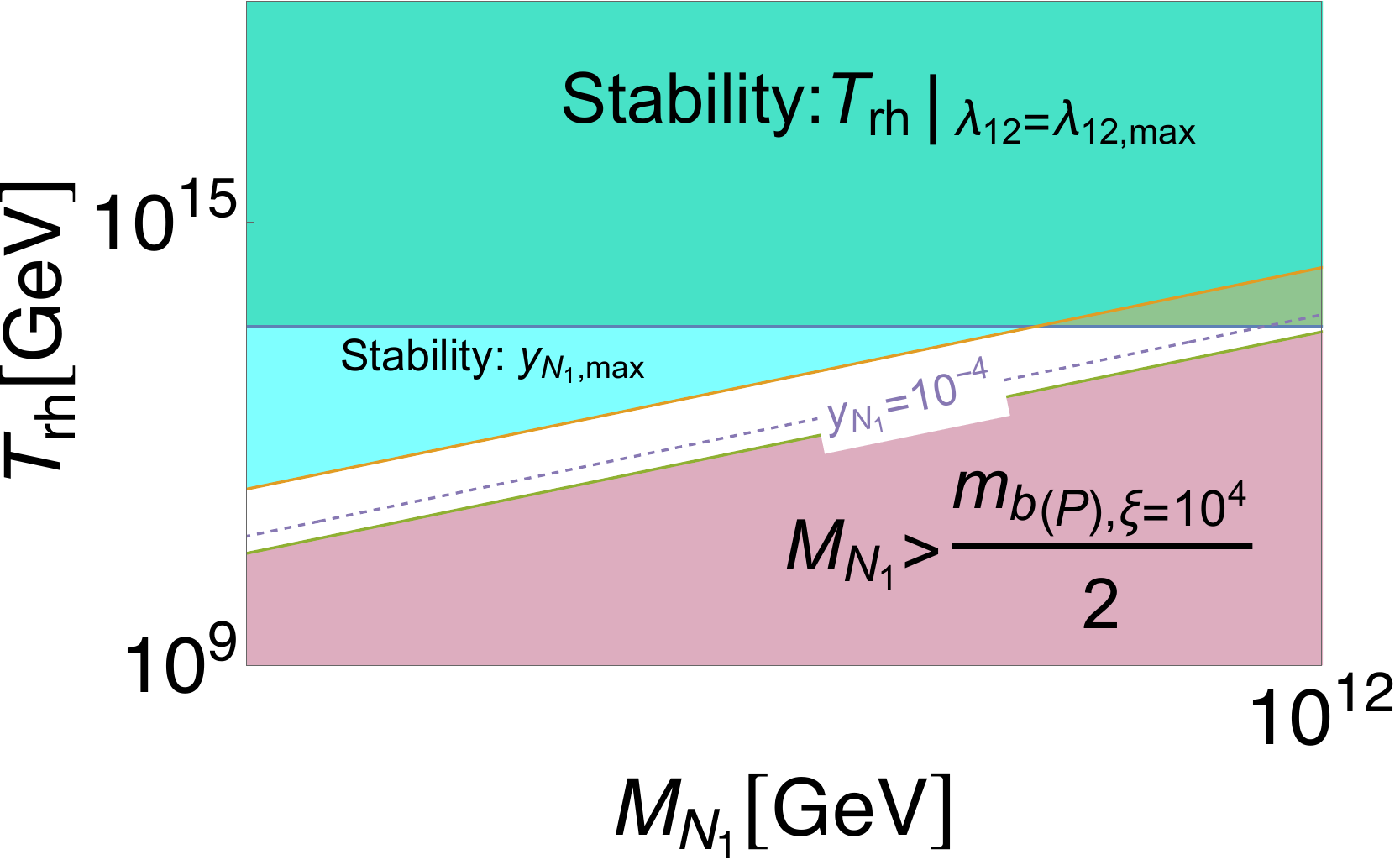}
    \caption{\it  \raggedright 
    The allowed region (unshaded area) on $(\Trh,M_{N_1})$ plane and viable range for the Yukawa-like coupling $y_{N_1}$ to produce the baryon asymmetry of the present universe for \nmpq~in Palatini formalism: {\bf top-left-panel} for the benchmark \Pmoon~for $\upzeta=1$(top-left), {\bf top-right-panel} is also for the benchmark \Pmoon~but for $\upzeta=10$, and {\bf bottom-panel} is for the benchmark \Psun~and for $\upzeta=10$.
    Thus, the parameter space of $y_{N_1}$ within Palatini formalism is tightly constrained.}
    \label{fig:pal-lepto-demo}
\end{figure}


However, Palatini formalism puts more stringent bounds on the production of sterile neutrinos as it can be seen from~\cref{fig:pal-lepto-demo}. The bounds on these figures are : 
    The maximum permissible value of $\Trh$ obtained from the stability analysis in~\cref{Table:Quartic_Palatini_stability} i.e. $\lt.\Trh\rt|_{\lambda_{12}=\lambda_{12,{\rm max}}}$. This is depicted as a forest-green horizontal stripe positioned at the top of the figures. The cyan-colored wedge-shaped area in the upper-left corner of the figures corresponds to the highest acceptable values of $y_{N_1}$ from~\cref{Table:Quartic_Palatini_stability}.
    The presence of a copper-rose colored region effectively excludes the possibility of the value of $M_{N_1}$ surpassing $ {\baremass}_{(P),\xi}/2$.
Top-left panel of~\cref{fig:pal-lepto-demo} is for the benchmark \Pmoon~with $\upzeta=1$. For this scenario, the possibility of the sterile neutrino generated from inflaton decay contributing to baryon asymmetry is feasible only for $y_{N_1}\sim 10^{-3.6}$.  Top-left panel of~\cref{fig:pal-lepto-demo}  is for the benchmark \Pmoon~with $\upzeta=10$, and this plot shows that the sterile neutrino from inflaton decay cannot account for the present-day baryon asymmetry within this scenario. However, the bottom panel reveals that successful Leptogenesis remains feasible for \Psun, even with $\upzeta=10$, if $y_{N_1}\sim 10^{-4}$.




\section{Discussion and Conclusion }
\label{Sec:Conclusion}
In this article, we considered slow-roll, single-field inflationary scenarios with quartic potential of the inflaton, and a non-minimal coupling between the inflaton and the gravity, and explored the particle production scenario from there-in. To this end, we suggested production of non-thermal fermionic \bsm~particle, which can be either stable or unstable, during the post-inflationary reheating era. 
Stable BSM fermionic particle can be accountable for the total CDM density of the universe. If the BSM fermionic particles are sterile neutrinos, they can decay to generate the total baryon asymmetry in the present universe via leptogenesis. Next, we studied a set of benchmark values for slow roll inflation satisfying bounds from CMB data within both metric and Palatini formalisms, and  explored the corresponding parameter space of the coupling between the \bsm~fermionic field and inflaton, and the mass of the fermionic field that leads to successful DM relic and leads to observed baryon asymmetric of the universe. The salient features of our analysis are as follows:

\begin{itemize}
    \item For our chosen benchmarks, the predicted values of $n_s$ and $r$ in both metric and Palatini formalisms, falls within $1-\sigma$ contour of combined analysis of Planck2018+Bicep3+Keck Array2018 on $(n_s,r)$ plane for $\ncmb\sim 60$ (see~\cref{fig:ns-r-contour}). For $\xi=0.5$, the predicted value of $r$ in Palatini formalism can be as small as $10^{-3}$, such that it can be tested in the future at $1-\sigma$ CL by forthcoming \cmbsfour~experiment. We also observed from~\cref{table:quartic-Benchmark-metric,table:quartic-Benchmark-Palatini}, as well as from~\cref{fig:ns-r-versus-Xi}, that the difference between predicted values of $r$  in Palatini formalism and in metric formalism, for the same $\xi$, increases for higher values of $\xi$. This is because $r$ varies as $\xi^{-1}$.  Left panel of~\cref{fig:ns-r-versus-Xi} shows that increasing the value of $\xi$ and considering Palatini formulation of gravity helps to rescue the inflationary model, even when the predicted values of $r$ in metric formalism are invalidated by CMB observations.

    


    \item From stability analysis, we obtained the upper limit of the coupling between the inflaton and the \bsm~fermionic field as $< O(10^{-4})$ in both metric and Palatini formalisms. However, the  upper limit of $\lambda_{12}$ obtained in metric formalism exceeds the value obtained in Palatini formalism by one order (see~\cref{Table:Quartic_metric_stability,Table:Quartic_Palatini_stability}). 

    \item Since $m_\phi=0$ in both metric and Palatini formalisms, we suggested incorporating a bare mass term to the potential for successful reheating. From our estimation, we found that the upper limit of the bare mass term in metric formalism depends on $\xi$ (\cref{eq:baremass-metric}), whereas in Palatini formalism, it depends on $\Trh$ (\cref{eq:bar mass limit Trh}).

    \item In~\cref{fig:Tmax-vs-Trh-plot}, we found that $\Tmax/\Trh$ is higher in metric formalism than in Palatini formalism for same values of $\xi$ and $\ncmb$. This is due to the fact that the inflaton potential around the minimum in Palatini formalism is quartic while it is quadratic around minimum in metric formalism. 
    We also observed that $\Tmax/\Trh$ differs noticeably with $\xi$ in Palatini formalism in contrast to metric formalism.

    \item From~\cref{fig:Dm yield plot 2 metric+Palatini+Palatini} we conclude that stable non-thermal $\chi$ produced only through inflaton decay can account for $100\%$ of the total relic density of CDM of the present universe, and the allowed range of $\yc$ is $10^{-3.5}\gsim \yc\gsim 10^{-20}$ in metric  formalism, while $10^{-4}\gsim \yc\gsim 10^{-10}$ in Palatini formalism.
    Unlike metric formalism, we observed that the permissible range of $y_{N_1}$ varies noticeably with $\xi$ in Palatini formalism. Furthermore, in contrast to metric formalism, the permissible upper limit of $\mc$ depends on $\Trh$.
The benchmarks chosen for $\ncmb~60$ in metric formalism (\cref{table:quartic-Benchmark-metric}) will be validated by future CMB observations, confirming the validity of the range of $\yc$ ($10^{-11}\gsim \yc \gsim 10^{-20}$). Additionally, CMB-S4 will affirm benchmarks in Palatini formalism, such as (P)BM3, indicating a valid range of $\yc$ around $10^{-11.5}$ for $\upzeta=10$. In summary, future CMB observations, such as \cmbsfour, can confirm given scenarios for dark matter physics and matter-antimatter asymmetry in metric and Palatini formalisms, with respect to BSM parameters $\yc$ and $\mc$.

    \item When sterile neutrinos are produced from the decay of inflaton and generates total baryon asymmetry of the universe,  we found from~\cref{fig:metric-lepto} that permissible range $y_{N_1}$ is $10^{-3.5}\gsim y_{N_1} \gsim 10^{-12}$ in metric formalism. However, Palatini formalism puts more stringent bounds on parameter space available for $y_{N_1}$.  
    For instance, in the case of \Pmoon, generating the baryon asymmetry necessitates $y_{N_1}\sim 10^{-3.5}$ when bare mass is approximately equal to its upper limit. But there is no permissible range of $y_{N_1}$ when the bare mass is one order below the upper limit, or even smaller. 
    This suggests that generating the universe's total baryon asymmetry through leptogenesis involving sterile neutrinos produced from the decay of inflaton is severely constrained for our chosen inflationary scenario in Palatini formalism.

\end{itemize}   
In summary, just by extending the standard model with two degrees of freedom: a real scalar inflaton, and a fermionic DM, we showed that tiny temperature fluctuations in CMBR can be explained via inflation and also addressed
the DM puzzle of the Universe. We have also made a comparative analysis between the metric and Palatini formalisms of gravity for the entire scenario. We have also demonstrated that future measurements of the CMB from experiments like CMB-S4, SPTpol, LiteBIRD, and CMB-Bharat~\cite{CMB-S4:2016ple,SPT:2019nip,LiteBIRD:2020khw,Adak:2021lbu} will further be able to test the simple models we have studied. 
We believe that, apart from producing necessary amount of DM and demonstrating successful Leptogenesis, starting from an well-motivated model of inflation in the two different gravity formalisms lead to different bounds on the parameter spaces thereby leaving significant effects on the corresponding physical conclusions of the theory. This is the major finding of the analysis, that makes it a compelling case to explore further. 

As a future outlook we would like to 
analyze the scalar perturbations and see if large density fluctuations at small scales can occur in these models leading to PBH production and scalar-induced Gravitational Waves in upcoming GW experiments like Laser Interferometer Space Antenna (LISA), Cosmic Explorer, Einstein Telescope~\cite{Amaro-Seoane:2012aqc,Evans:2021gyd,Maggiore:2019uih}. we leave such estimates for future studies.

\medskip

\section*{Acknowledgement}
Work of Shiladitya Porey is funded by RSF Grant 19-42-02004. Supratik Pal thanks Department of Science and Technology, Govt. of India
for partial support through Grant No. NMICPS/006/MD/2020-21.

\appendix

\section{CW inflation}
%
%

\begin{figure}[H]
    \centering
    \includegraphics[width=0.5\linewidth]{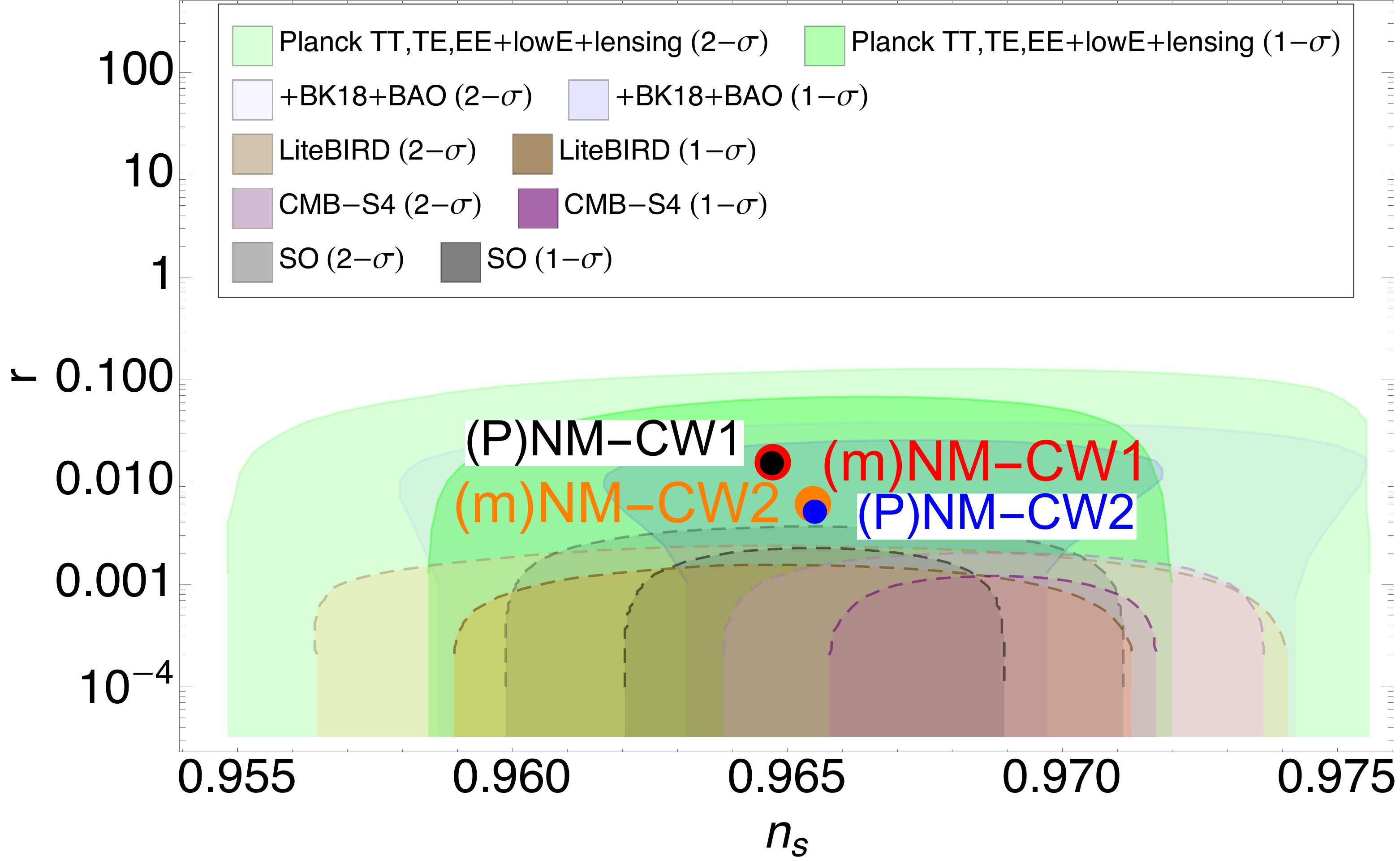}
    \caption{\it \raggedright
    $(n_s-r)$ predictions for $\xi=-0.002$ and $\xi=-0.0025$ in both metric and Palatini formalism and for $ \ncmb\sim 60$ (from~\cref{Table:Benchmark-CW-metric,Table:Benchmark-CW-Palatini}) along with bounds on  $(n_s-r)$ from current and future prospective reaches from forthcoming \cmb~observations mentioned in~\cref{fig:ns-r-contour}.
    }
    \label{fig:CW_ns-r}
\end{figure}

Since, we consider a \sm~gauge singlet \bsm~field as the inflaton with quartic potential in our work, the presence of self-interaction or Yukawa interaction in the Lagrangian density can give rise to quantum corrections. When such quantum loop corrections originating from interaction terms, UV completions~\cite{Racioppi:2018zoy}, etc., are included, the effective potential for inflaton in \jframe~becomes
\begin{align}\label{Eq:CW_potential_JordanFrame}
V^{JF}(\vp)= V_{\nmcw}(\varphi)= \frac{\Lambda \cq^4}{4}+\Lambda \varphi ^4 \left[\log \left(\frac{\varphi }{\cq}\right)-\frac{1}{4}\right]\,,
\end{align}
where $\cq$ represents the renormalization scale, and the inclusion of the term $\frac{\Lambda \cq^4}{4}$ is to ensure that the value of the potential is $\gsim 0$, even at its minimum (the minimum of $V_{\nmcw}(\varphi)$ is located at $\vp=\cq$).  With an aim to find benchmark values for which predicted values of $(n_s,r)$ fall within $1-\sigma$ best fit contour of \Planck2018+\BICEP3+\KeckArray2015 combined analysis, unlike the form of non-minimal coupling considered in~\cref{eq:form of OmegaSq}, we consider the non-minimal coupling of the following form~\cite{Maji:2022jzu,Ghoshal:2023jhh}
\be \label{eq:shafi-non-minimal-coupling}
\Omega^2(\vp)= 1+ \xi \frac{\vp^2- \cq^2}{\mpl^2}\,. 
\ee 
This particular form of non-minimal coupling also ensures that $\Omega^2(\vp\to \cq)\to 1$. 
Consequently, \eframe~and \jframe~becomes the same when inflaton reaches to the minimum of the potential. 
The potential for non-minimal Coleman Weinberg slow roll inflation (\nmcw) in \eframe~is
\eq{\label{Eq:PotCW-Einsteinframe}
V^E(\vp)\equiv V_{\nmcw}^E(\vp)
= \frac{    \frac{\Lambda \cq^4}{4}+\Lambda \varphi ^4 \left[\log \left(\frac{\varphi }{\cq}\right)-\frac{1}{4}\right]   }{\lt( 1+ \xi \frac{\vp^2-\cq^2}{\mpl^2}\rt)^2}\,.
}
For the form of non-minimal coupling in~\cref{eq:shafi-non-minimal-coupling}, the relation between the inflaton in \eframe~and $\vp$ is 
\eq{\label{Eq:JordantoEinsteinNew}
\dv{\phi}{\vp}=
\bc
\sqrt{
\frac{\xi  (6 \xi +1) \varphi^2 M_P^2 - \xi \cq^2\mpl^2+M_P^4}{\left(M_P^2+\xi  (\varphi
   ^2-\cq^2)\right){}^2}
}  
& \text{(in metric theory)}\,,
\\
\sqrt{
\frac{\mpl^2}{\left(M_P^2+\xi  (\varphi
   ^2-\cq^2)\right)}
}
& \text{(in Palatini theory)}\,.
\ec 
}
A set of benchmark values for slow roll inflation with potential $V_{\nmcw}^E(\vp)$ in \eframe~in metric and Palatini formalism are mentioned for $\ncmb=50$ and $\ncmb=60$ in~\cref{Table:Benchmark-CW-metric,Table:Benchmark-CW-Palatini}. From those tables, we see that $\abs{\xi}<1$ for our chosen benchmark values. In this scenario, slow roll inflation  initiates near $\phi=0$ and inflaton moves towards higher value of $\phi$ as inflation progresses. The picture is opposite for slow roll inflation in plain quartic potential for inflaton, where the inflaton moves from larger to smaller values of inflaton during slow roll inflation. 
The predicted values of $(n_s, r)$ fall within $1-\sigma$ best-fit contour of \Planck+\BICEP+\KeckArray~combined data for $\ncmb=60$, which is similar to inflation with plain quartic potential of inflaton (see~\cref{table:quartic-Benchmark-metric,table:quartic-Benchmark-Palatini}). Therefore, hereafter, we consider only the benchmark values corresponding to $\ncmb=60$ for further analysis. The predicted $(n_s,r)$ values for benchmark (m)NM-CW1, (m)NM-CW3, (P)NM-CW1, and (P)NM-CW3 are shown in~\cref{fig:CW_ns-r} along with present $1-\sigma$ and $2-\sigma$ contour from \Planck~and \Planck+\BICEP+\KeckArray~mission, along with prospective future reaches  from forthcoming \cmb~observations as mentioned in~\cref{fig:ns-r-contour}. From this figure, we can see that for the same values of $\xi$ and $\ncmb$ for our chosen set of benchmarks, there is no large difference between the predicted values of $(n_s,r)$ in metric and Palatini formalisms. This is in contrast to what we observed for a simple quartic potential for inflation, as illustrated in~\cref{fig:ns-r-contour}. The reason for these almost similar predicted values of $(n_s,r)$ is our choice of $\abs{\xi}<1$. Additionally, the chosen four benchmarks ((m)NM-CW1, (m)NM-CW3, (P)NM-CW1, and (P)NM-CW3) for $\ncmb\sim 60$, can be confirmed in the future with the help of the three forthcoming \cmb~observations, the prospective future reaches of which are illustrated in~\cref{fig:CW_ns-r}. 
%
\begin{table}[H]
\centering
\caption{\it Benchmark values for \nmcw~inflationary model in metric formalism.}
\label{Table:Benchmark-CW-metric}
\begin{tabular}{|c||c | c | c | c | c | c | c| c| } 
 \hline
{\it Benchmark}&$\cq/\mpl$ &$\xi$& $\vp_{\rm end}/\mpl$ &  $\ncmb$& $\vp_{*}/\mpl$  &  $n_s$ & $r\times 10^{2}$ & $\L$ \\ %
 \hline 
 \hline 
  (m)NM-CW1  & \multirow{2}{*}{$52$} & \multirow{2}{*}{$-0.0020$}& \multirow{2}{*}{$ 50.7913$} & $60$ &$39.0185$  & $0.9647$
  & $1.5708$ & $9.4735\times 10^{-15}$ \\
 \cline{1-1}\cline{5-9}
  (m)NM-CW2  &  & & & $50$  & $39.8917$  & $0.9595$ & $2.2375$ & $  1.3815\times 10^{-14}$ \\
 \hline
  (m)NM-CW3  & \multirow{2}{*}{$175$} & \multirow{2}{*}{$-0.0025$}& \multirow{2}{*}{$ 174.1685$}& $60$  & $161.9637$  & $0.9655$
  & $0.6114$ & $ 2.9743\times 10^{-15}$ \\
 \cline{1-1}\cline{5-9}
  (m)NM-CW4  &  & &  & $50$ & $162.9510$  & $0.9603$ & $0.8765$ & $ \  4.3037\times 10^{-15}$ \\
 \hline
\end{tabular}
\end{table}
%

 %
\begin{table}[H]
\centering
\caption{\it Benchmark values for \nmcw~inflationary model in Palatini formalism.}
\label{Table:Benchmark-CW-Palatini}
\begin{tabular}{|c||c | c | c | c | c | c | c| c| } 
 \hline
{\it Benchmark}&$\cq/\mpl$ &$\xi$& $\vp_{\rm end}/\mpl$ &  $\ncmb$& $\vp_{*}/\mpl$  &  $n_s$ & $r\times 10^{2}$ & $\L$ \\ %
 \hline 
 \hline 
  (P)NM-CW1  & \multirow{2}{*}{$52$} & \multirow{2}{*}{$-0.0020$}& \multirow{2}{*}{$ 50.7649$} & $60$ &$38.9338$  & $0.9647$
  & $1.5332$ & $9.2276\times 10^{-15}$ \\
 \cline{1-1}\cline{5-9}
  (P)NM-CW2  &  & & & $50$  & $39.8033$  & $0.9595$ & $2.1851$ & $  1.3456\times 10^{-14}$ \\
 \hline
  (P)NM-CW3  & \multirow{2}{*}{$175$} & \multirow{2}{*}{$-0.0025$}& \multirow{2}{*}{$ 173.9785$}& $60$  & $161.2729$  & $0.9655$
  & $0.5104$ & $ 2.4699\times 10^{-15}$ \\
 \cline{1-1}\cline{5-9}
  (P)NM-CW4  &  & &  & $50$ & $162.2357$  & $0.9605$ & $0.7321$ & $3.5699\times 10^{-15}$ \\
 \hline
\end{tabular}
\end{table}
%

%
%
%
%

%
%
\begin{table}[H]
\begin{center}
\caption{\it Allowed range of $\yc$ and $\lO$ for the benchmark values from~\cref{Table:Benchmark-CW-metric}.}
\label{Table:NM-CW_stability-metric}
\vspace{-8pt}
\resizebox{\columnwidth}{!}{
\begin{tabular}{|c|| c| c| c|c|} 
 \hline
 {\it Benchmark} &\multicolumn{2}{|c|}{stability for $\yc$} & \multicolumn{2}{|c|}{stability for $\lO$}\\ [0.5ex] 
  \cline{2-5} 
   & about $\phi_*$ & about $\phi_{\rm end}$ & about $\phi_*$ & about $\phi_{\rm end}$ \\ [0.5ex] 
 \hline\hline
(m)NM-CW1 & $\yc<3.0956\times 10^{-4}$ & $\yc<3.0671\times 10^{-4}$ & $\lO/\mpl<2.3304\times 10^{-6}$ & $\lO/\mpl<3.3615\times 10^{-6}$ \\ 
 \hline
(m)NM-CW3 & $\yc< 1.6355\times 10^{-4}$ & $\yc< 2.5870\times 10^{-4}$ & $\lO/\mpl<8.9651 \times 10^{-7}$ & $\lO/\mpl< 2.7832\times 10^{-6}$ \\
 \hline
\end{tabular}
}
\end{center}
\end{table}


%
%
\begin{table}[H]
\begin{center}
\caption{\it Allowed range of $\yc$ and $\lO$ for the benchmark values from~\cref{Table:Benchmark-CW-Palatini}.}
\label{Table:NM-CW_stability-Palatini}
\vspace{-8pt}
\resizebox{\columnwidth}{!}{
\begin{tabular}{|c|| c| c| c|c|} 
 \hline
 {\it Benchmark} &\multicolumn{2}{|c|}{stability for $\yc$} & \multicolumn{2}{|c|}{stability for $\lO$}\\ [0.5ex] 
  \cline{2-5} 
   & about $\phi_*$ & about $\phi_{\rm end}$ & about $\phi_*$ & about $\phi_{\rm end}$ \\ [0.5ex] 
 \hline\hline
(P)NM-CW1 & $\yc<3.0724\times 10^{-4}$ & $\yc<3.0673\times 10^{-4}$ & $\lO/\mpl<2.2859\times 10^{-6}$ & $\lO/\mpl<3.3423\times 10^{-6}$ \\ 
 \hline
(P)NM-CW3 & $\yc< 1.5297\times 10^{-4}$ & $\yc< 2.5183\times 10^{-4}$ & $\lO/\mpl<7.7185 \times 10^{-7}$ & $\lO/\mpl< 2.5579\times 10^{-6}$ \\
 \hline
\end{tabular}
}
\end{center}
\end{table}


Following the similar approach used in~\cref{sec:stability}, we estimated permissible upper values of $\yc$ and $\lO$ for the interaction Lagrangian~\cref{Eq:reheating lagrangian} and for $V^E_{\rm tree}\equiv V_{\nmcw}^E(\vp)$, and it is mentioned in~\cref{Table:NM-CW_stability-metric} (for metric formalism) and in~\cref{Table:NM-CW_stability-Palatini} (for Palatini formalism). From these table, we conclude that upper permissible values of $\yc$ and $\lO$ are
\begin{itemize}
    \item for (m)NM-CW1: $\yc<3.0671\times 10^{-4}$, $\lO/\mpl<2.3304\times 10^{-6}$.
    \item for (m)NM-CW3: $\yc<1.6355\times 10^{-4}$, $\lO/\mpl<8.9651 \times 10^{-7}$.
 \item  for (P)NM-CW1: $\yc<3.0673\times 10^{-4}$, $\lO/\mpl<2.2859\times 10^{-6}$.
    \item for (P)NM-CW3: $\yc<1.5297\times 10^{-4}$, $\lO/\mpl<7.7185 \times 10^{-7}$.  
\end{itemize}
These upper permissible values of $\lO$ leads to the estimation of the maximum permissible value of $\Trh$. We make the assumption that for \nmcw~inflation, reheating happens in quadratic potential for both metric and Palatini formalisms. Furthermore, $\vp_{\rm min}=\cq$ is the location of minimum  of $V_{\nmcw}^E(\vp)$ and $m_\phi=\qty(\left.\qty({\dd^2 V^E}/{\dd \phi^2}) \right|_{\varphi=\varphi_{\rm min}})^{1/2}$ is not vanishing, as we have seen for plain quartic inflation (see~\cref{Table:CW-mass-inflaton}). 
Hence, unlike \nmpq, we don't require introducing a bare mass term to the potential. The assumption that potential in \eframe~around the minimum in \nmcw~inflation can be approximated as $\propto\phi^2$, even in Palatini formalism, along with non-vanishing values of  $m_\phi$, leads to the different results in the context of DM production and leptogenesis, compared to \nmpq,  that we obtain next for the production of DM and Leptogenesis.

 %
\begin{table}[H]
\centering
\caption{\it $m_\phi$ for \nmcw~inflationary model in both metric and Palatini formalism.}
\label{Table:CW-mass-inflaton}
\begin{tabular}{|c||c |   } 
 \hline
 {\it Benchmark} & $m_\phi/\mpl$\\
 \hline
 \hline 
 (m)NM-CW1 & $9.8092\times 10^{-6}$\\
 \hline 
  (m)NM-CW3 & $1.3023\times 10^{-5}$\\
 \hline 
  (P)NM-CW1 & $9.9903\times 10^{-6}$\\
 \hline 
  (P)NM-CW3 & $1.7395\times 10^{-5}$\\
 \hline 
\end{tabular}
\end{table}
 %

 %
\begin{figure}[H]
    \centering
    \includegraphics[width=0.45\linewidth]{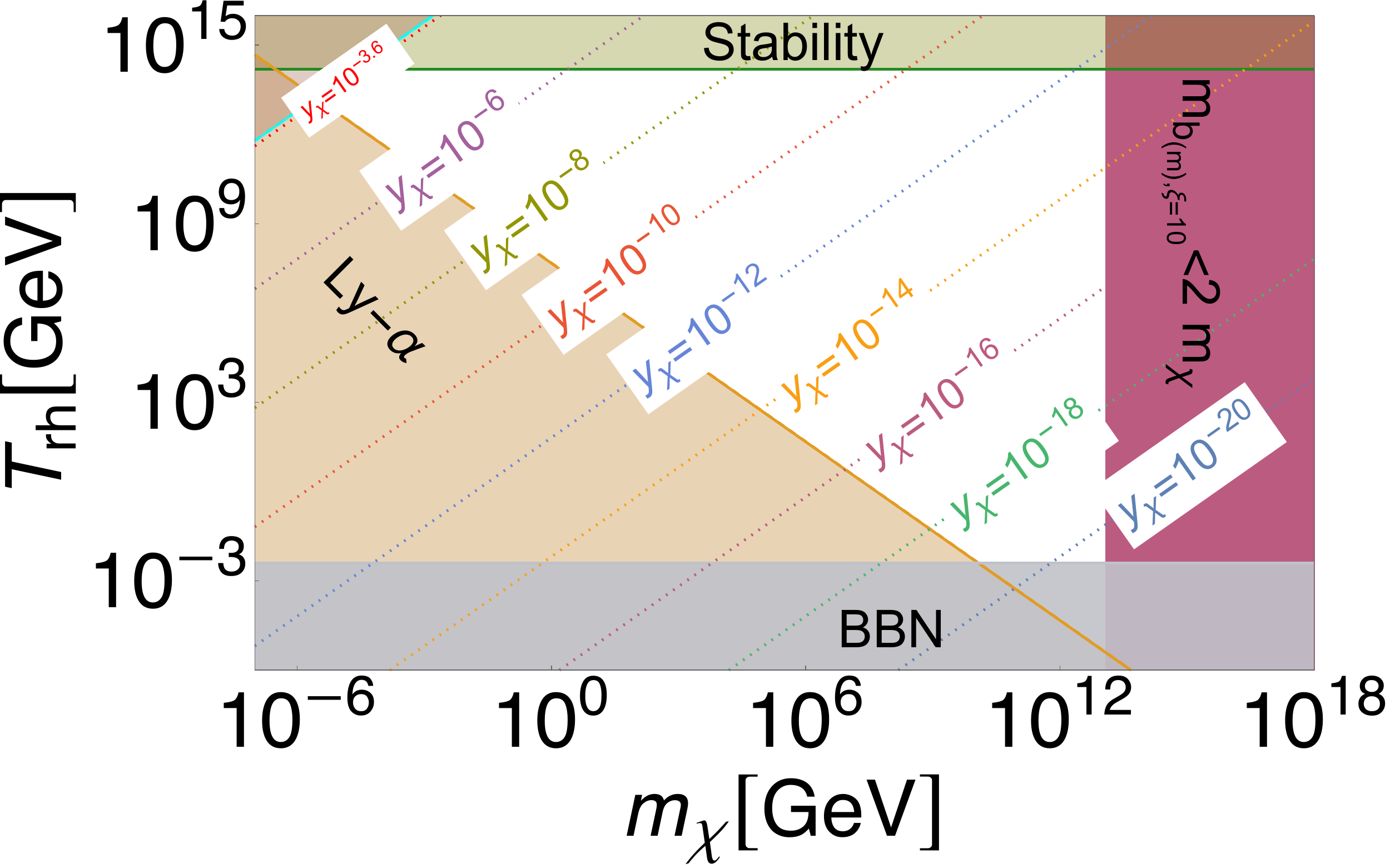}\;
    \includegraphics[width=0.45\linewidth]{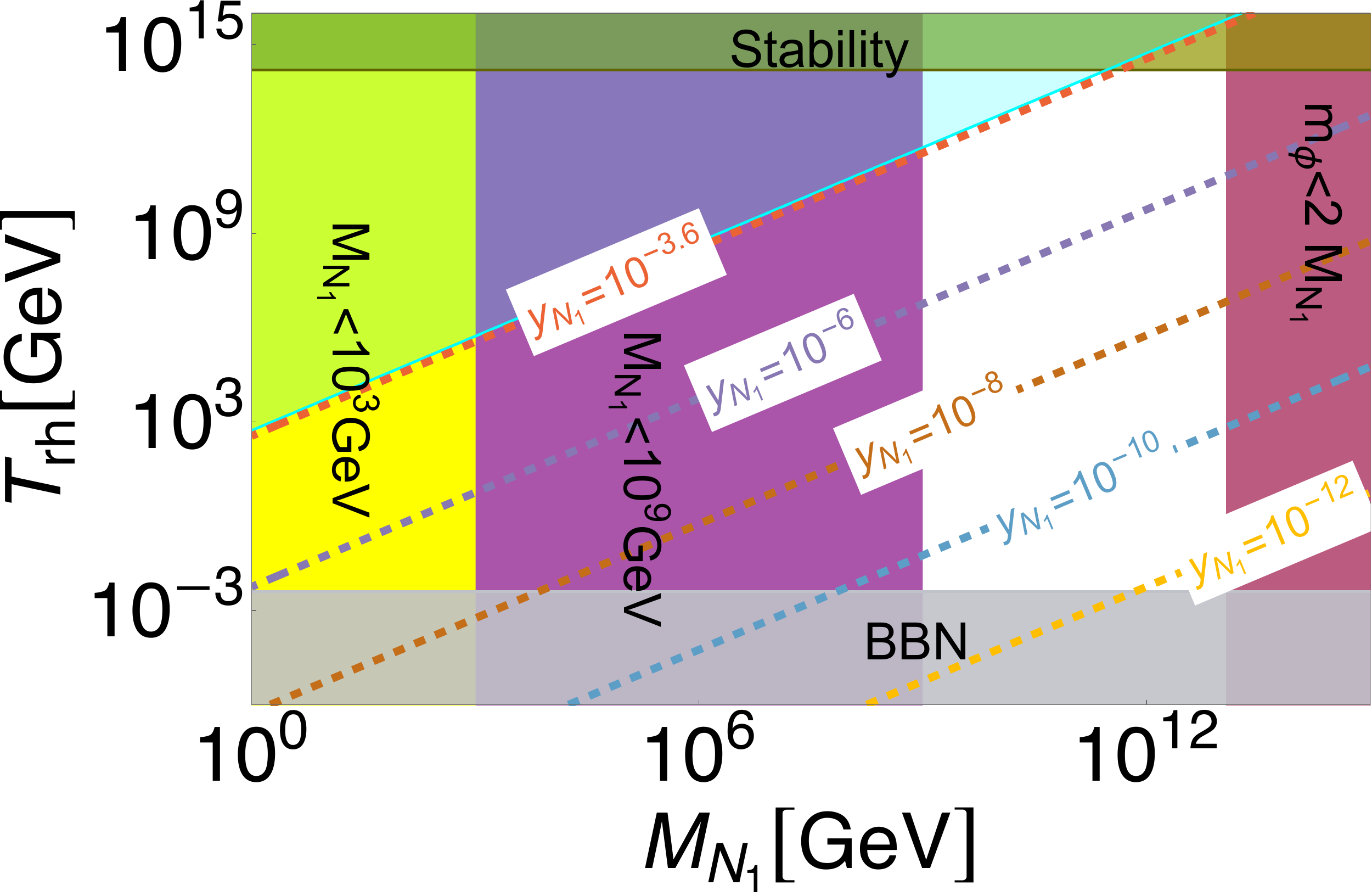}
    \caption{\it \raggedright
    {\bf Left panel:} shows
    the allowed region (unshaded area) on $(\Trh,\mc)$ plane and viable range for the Yukawa-like coupling $\yc$ to produce the entire \cdm~density of the present universe. {\bf Right panel:} displays the allowed region (unshaded area) on $(\Trh,M_{N_1})$ plane and viable range for the Yukawa-like coupling $y_{N_1}$ to produce the baryon asymmetry of the present universe. Both panels are for \nmcw~in metric formalism for the benchmark (m)NM-CW1. The allowed regions on $(\Trh,\mc)$ plane and viable range for $\yc$, and allowed regions on $(\Trh,M_{N_1})$ plane and viable range of $y_{N_1}$ are almost identical for the benchmark (P)NM-CW1. This similarity arises from the almost identical values of $m_\phi$ and permissible upper limits of $\yc$ and $\lO$, with slight variations, for both benchmarks (m)NM-CW1 and (P)NM-CW1.
    }
    \label{fig:NMCW+DM+Lepto}
\end{figure}
%
Left panel of~\cref{fig:NMCW+DM+Lepto} is similar to~\cref{fig:Dm yield plot 2 metric+Palatini+Palatini}, but for \nmcw~(benchmark - (m)NM-CW1). In this figure, the dashed lines correspond to~\cref{eq:Yield of DM-metric} illustrating the values of $\yc$ for which $\chi$ produced from the decay of inflaton can be accountable for the $100\%$ of total \cdm~density of the present universe. The bounds on $(\Trh,\mc)$ plane in~\cref{fig:NMCW+DM+Lepto} have been already mentioned in~\cref{fig:Dm yield plot 2 metric+Palatini+Palatini}: from stability analysis mentioned in~\cref{Table:NM-CW_stability-metric}, $m_\chi$ should be $\lsim m_\phi/2$ ($m_\phi$ is mentioned in~\cref{Table:CW-mass-inflaton}), Ly-$\alpha$ bound from~\cref{eq:Lyman-alpha}, \bbn~bound: $\Trh\gsim 4 \MeV$. The allowed range of $\yc$ is $10^{-3.6}\gsim \yc \gsim 10^{-20}$, which is identical for metric formalism in \nmpq. In \nmcw, the allowed range of $\yc$ is also nearly the same for Palatini formalism (benchmark (P)NM-CW1), because $m_\phi$, and the upper permissible limit of $\yc$ and $\lO$ are almost identical in both metric and Palatini formalisms for our chosen benchmark values. Right panel of~\cref{fig:NMCW+DM+Lepto} is similar to~\cref{fig:metric-lepto}, but for \nmcw~(benchmark - (m)NM-CW1) depicting allowed range for $y_{N_1}$ by dotted lines. The unshaded region is allowed, and the bound on $(\Trh, M_{N_1})$ plane have been already mentioned in~\cref{fig:metric-lepto}. The allowed range of $y_{N_1}$ is $10^{-3.6}\gsim y_{N_1} \gsim 10^{-12}$, which is identical for metric formalism in \nmpq.  The allowed region in Palatini formalism is identical and this conclusion is different from the result we obtained for \nmpq.

\section{Boltzmann}
\label{appendix:Boltzmann}
Here, we derive the expression for $\Tmax$ while reheating happens in either quadratic or quartic potential about the minimum. However, to make the calculations more general, we assume the potential during reheating $V^E\propto \phi^k$ where $k$ can be any even number $>0$ so that we can get a proper minimum during reheating. 
Then, the Boltzmann equations governing the evolution of $\r_\phi$, $\rr$, and $\r_\chi$ (which refers to energy density of $\chi$), and first Friedmann equation are given by~\cite{Bernal:2019mhf,Giudice:2000ex,Garcia:2020eof}%
\footnote{In contrast to our approach or the approach used in~\Ccite{Bernal:2019mhf},~\Ccite{Garcia:2020eof} considers that $m_\phi$ is not a constant parameter. As a result, there is a different relationship between $\rho_{\rad}$ and $\cs$ and $T$ as compared to the one we derive below.}
\eq{
&\frac{\td \r_\phi}{\td t} 
+ 3 \left( \frac{2k}{k+2}\right)
\hubble \rho_\phi = - \Gamma_\phi \rho_\phi.
\label{Eq:B-rhophi} \\
&
\frac{\td \rho_{\rm rad}}{\td t}  + 4 \hubble \rho_{\rm rad} \;=\; \Gamma_\phi \rho_\phi\,,
\label{Eq:B-rhorad}
\\
&\frac{\td n_\chi}{\td t} + 3\hubble n_\chi \;=\; R(t)\,,\label{Eq:B-nchi}\\
&
\hubble^2 \;=\; \frac{\rho_\phi + \rho_{\rm rad} +\rho_\chi}{3 \mpl^2} \;\approx\; \frac{\rho_{\phi}}{3 M_P^2}\, ,
\label{Eq:B-Hubble}
}
The approximation presented in~\cref{Eq:B-Hubble} holds true starting from the onset of reheating era and prior to the point when the temperature of \sm~relativistic plasma, or equivalently, the temperature of the universe reaches $\Trh$. When $\Gamma_\phi\ll \hubble$, which occurs during the early stages of reheating, we can approximate the right-hand side of~\cref{Eq:B-rhophi} as $0$. Under this approximation, the solution to~\cref{Eq:B-rhophi} can be written as $\rho_\phi \propto \cs^{-3\lt( 2 k/ (k+2) \rt)}$. 
To further simplify the calculations of combined~\cref{Eq:B-rhophi,Eq:B-rhorad,Eq:B-nchi,Eq:B-Hubble}, we make use of comoving energy density $\rho_\phi= R_\phi \, \cs^{-3\lt( 2 k/ (k+2) \rt)}$, $\r_{\rm rad} \sim R_{\rm rad} \, \cs^{-4}$, $\r_\chi = R_\chi\, \cs^{-3}$. By combining them with~\cref{Eq:B-rhorad,Eq:B-Hubble}, we obtain the following 
\eq{
\frac{\td R_{\rm rad}}{\td a}  \;=\; \sqrt{3}\mpl\, \Gamma_\phi \sqrt{R_{\phi}} \, \cs^{6/(2+k)}\,,
}
which is applicable at early times i.e. when $\hubble \gg \Gamma_\phi$. This allows us to assume $R_{\phi} \approx R_{\phi_I} $, where $R_{\phi_I}$ is the value of $R_{\phi}$ at the beginning of reheating and $\frac{\td R_{\phi_I}}{\td \cs}\approx 0$. Then 
\eq{
R_{\rm rad} \simeq \sqrt{3}\mpl\,\Gamma_\phi\,\sqrt{R_{\phi_I}}\, \frac{2+k}{8+k} \cs^{(8+k)/(2+k)} + \mathbb{c}\,,
}
where $\mathbb{c}$ is the indefinite integral constant that may be settled with the assumption that
\eq{
R_{\rm rad} (\cs_I) \simeq 0\,,
}
Therefore,
\eq{\label{Eq:B-rad-2}
R_{\rm rad} \simeq \sqrt{3}\mpl\,\Gamma_\phi\,\sqrt{R_{\phi_I}}\, \frac{2+k}{8+k} \lt(\cs^{(8+k)/(2+k)} -\cs_I^{(8+k)/(2+k)}\rt)\,,
}
To further facilitate the computations, let us introduce a variable $\csn\equiv \cs/\cs_I$. 
Besides, energy density of the produced relativistic \sm~particles which are thermalized and posses the temperature $T$ can be written as 
\eq{\label{Eq:energy density of radiation}
R_{\rm rad}=\frac{\pi^2}{30} \gs T^4 \csn^4 \cs_I^4\,.
}
Equating~\cref{Eq:B-rad-2,Eq:energy density of radiation}, we obtain
\eq{\label{Eq:expresion for Temp}
T=3^{3/8}\lt( \frac{10 \, \mpl\, \Gamma_\phi}{\pi^2\,\gs} \sqrt{R_{\phi_I}}\rt)^{1/4}
\qty(\frac{2+k}{8+k})^{1/4} 
\cs_I^{\frac{-3k}{4(2+k)}} 
\lt(\csn^{\frac{-3k}{(2+k)} }- \csn^{-4}
\rt)^{1/4}\,.
} 
In~\cref{Eq:expresion for Temp}, the maximum value of $T$ can be obtained if $\lt(\csn^{\frac{-3k}{(2+k)} }- \csn^{-4}\rt)^{1/4}$ reaches maximum, which happens when
\eq{
\csn_{\rm max}=\left(\frac{4}{3}
   \frac{k+2}{k}\right)^{\frac{k+2}{k+8}}
=\left(\frac{4}{3} \frac{2}{w+1}\right)^\frac{2}{5-3 w_\re}\,.
}

Then, the maximum possible temperature during reheating 
\eq{
\Tmax &=\left(\frac{60}{\pi^2\, \gs}\right)^{1/4}\, \mpl^{1/4}\, \Gamma_\phi^{1/4}\, R_{\phi_I}^{1/8}\,
\cs_I^{-\frac{3k}{4(k+2)}}\, \frac{3^{1/8}}{2^{3/4}}
 \lt(\left(\frac{27}{64}\right)^{\frac{k}{k+8}}
   \left(\frac{k}{k+2}\right)^{\frac{3 k}{k+8}}
   \rt)^{1/4}\,.\\
%
%
&=\left(\frac{60}{\pi^2\, \gs}\right)^{1/4}\, \mpl^{1/2}\, \Gamma_\phi^{1/4}\, {\hubble_I}^{1/4}\, \frac{3^{5/8}}{2^{3/4}}
 \lt(\left(\frac{27}{64}\right)^{\frac{k}{k+8}}
   \left(\frac{k}{k+2}\right)^{\frac{3 k}{k+8}}
   \rt)^{1/4}
}

Therefore,
\begin{empheq}[
  left=
    {
\Tmax=
} 
    \empheqlbrace
]{align}
    &\qty(\frac38)^{2/5} \, \lt( \frac{60}{\gs \, \pi^2}\rt)^{1/4} \qty(\hubble_I\, \Gamma_\phi)^{1/4} \, \mpl^{1/2} \, \qquad \text{(for quadratic)} \,,\label{eq:TMAX-quadratic}
    \\
    &\qty(\frac{3}{16})^{1/4} \, \lt( \frac{60}{\gs \, \pi^2}\rt)^{1/4} \qty(\hubble_I\, \Gamma_\phi)^{1/4} \, \mpl^{1/2} \, \qquad \text{(for quartic)}\,. \label{eq:TMAX-quartic}
\end{empheq}

Since $\csn^{-1}<\csn^{\frac{-3k}{4(2+k)}}$ for $\csn>1$ and $k>1$, we can approximate~\cref{Eq:expresion for Temp} as
\eq{\label{eq:Last pole}
T\approx 3^{3/8}\lt( \frac{10 \, \mpl\, \Gamma_\phi}{\pi^2 \, \gs} \sqrt{R_{\phi_I}}\rt)^{1/4}
\qty(\frac{2+k}{8+k})^{1/4} 
\cs_I^{\frac{-3k}{4(2+k)}} \,
\csn^{\frac{-3k}{4(2+k)} }\,.
} 

From~\cref{eq:Last pole}, we can see that $T \propto \cs^{{-3k}/{4(2+k)}}$. Additionally, using $\rho_\phi= R_\phi \, \cs^{-3\lt( 2 k/ (k+2) \rt)}$, we see that $\rho_\phi\sim T^8$ and this functional dependence remains unchanged regardless of whether we consider a quadratic or quartic potential for the inflaton during reheating.

\section{Upper limit estimation for bare mass}
\label{appendix:bare mass}
In this section, we estimate the upper limit on the bare mass that can be incorporated into the potential of inflaton for successful reheating without disrupting the slow roll inflationary scenario. 
We assume that 
the potential of inflaton around its minimum during reheating (in Einstein frame, with $\phi$ as the inflaton in Einstein frame) can be approximated using one of the following forms:
\eq{\label{eq:pot-reheat-form}
\tilde{V}^E(\phi) \propto\begin{cases}
    \Lambda_{\qud} \, 
    \phi^2\,,\\
    \Lambda_\qur \,  
    \phi^4\,,
\end{cases} 
}
where $ \Lambda_{\qud},\Lambda_{\qur}$ are numerical factors. 
Our initial goal is to calculate the upper limit for the bare mass when reheating occurrs at the quartic minimum.
By using~\cref{eq:w-k-relation} along with the relation $\r_\phi\propto \cs^{-3\qty(1+w_\re)}$, we obtain during the reheating era~\cite{Garcia:2020eof} 
\eq{\label{eq:rho-phi-cs-relation}
\r_\phi \propto
\cs^{-4}\,. 
}
Comparing~\cref{eq:pot-reheat-form,eq:rho-phi-cs-relation}, we can say that amplitude of oscillation varies as (for quartic potential)
\eq{\label{Eq:Inflaton-scale-factor}
\phi   \sim
\cs^{-1} \,.
}

Therefore, $m_\phi   \sim \cs^{-1}$ (as $m_\phi\sim \sqrt{\Lambda_\qur}\,\phi$), and $\Gamma_\phi \sim \cs^{1}$. Additionally, $\hubble \sim \cs^{-2}$ as $\hubble\sim \sqrt{\tilde{V}^E(\phi)}$.

Hence,
\eq{
\frac{\Gamma_\phi}{\hubble} \propto  
\cs^3 \,.
}
Therefore, using $\Gamma_{\phi,\rh} \sim \hubble_\rh$, we get
\eq{\label{eq:111}
\frac{\cs_\rh}{\cs_{\rm end}}  \sim 
%
\lt( \frac{\hubble_{\rm end}}{\Gamma_{\phi, {\rm end}}}  \rt)^{1/3}\,.
}
By using~\cref{Eq:Inflaton-scale-factor} in~\cref{eq:111}, we obtain
\eq{
\phi_\rh \sim 
%
\phi_{\rm end} 
\lt( \frac{\Gamma_{\phi,end}}{\hubble_{\rm end}} \rt)^{1/3}\,.
}
Then, using $\hubble_{\rm end} \sim \sqrt{V(\phi_{\rm end})}/(\sqrt{3}\mpl)$, 
\eq{
\phi_\rh  \sim 
%
%
\lt(\frac{\sqrt{3}\, \mpl}{\sqrt{\Lambda_\qur}} \Gamma_{\phi,{\rm end}} \,  \phi_{\rm end}\rt)^{1/3} \sim \lt(\frac{\sqrt{3}\, \mpl\, \lambda_{12}^2}{8\pi \,{\Lambda_\qur}} \rt)^{1/3}\,,
%
%
}
where we have used $\Gamma_{\phi,{\rm end}} = \frac{\lambda_{12}^2}{8 \pi \sqrt{\Lambda_\qur} \phi_{\rm end}}$. If we want to incorporate a bare mass term ${\baremass}$ during quartic reheating, we need to satisfy the following condition~\cite{Dimopoulos:2017xox}
\eq{\label{eq:appendix-bare-mass-upper-limit-Palatini}
{\baremass} &< \sqrt{\Lambda_\qur} \, \phi_\rh =\sqrt{\Lambda_\qur} \lt(\frac{\sqrt{3}\, \mpl\, \lambda_{12}^2}{8\pi \,{\Lambda_\qur}} \rt)^{1/3}\,.
}

For reheating happens in quadratic potential, the upper limit on bare mass can straightforwardly be estimated as 
\begin{align}\label{eq:appendix-bare-mass-upper-limit-metric}
    {\baremass} < \sqrt{\Lambda_\qud}\,.
\end{align}

\bibliographystyle{apsrev4-1}
\bibliography{reference}
\end{document}